\documentclass[article]{ajs}

\author{Adam Bilchouris\,\orcidlink{0009-0002-4649-0247}\\ La Trobe University,\\ Melbourne, Australia \\ \And 
        Andriy Olenko\,\orcidlink{0000-0002-0917-7000}\\ La Trobe University,\\ Melbourne, Australia}
\title{On Nonparametric Estimation of Covariograms}

\Plainauthor{Adam Bilchouris, Andriy Olenko} %% comma-separated
\Plaintitle{On Nonparametric Estimation of Covariograms} %% without formatting
\Shorttitle{Estimation of Covariograms} %% a short title (if necessary)

\Abstract{
The paper overviews and investigates several nonparametric methods of estimating covariograms. It provides a unified approach and notation to compare the main approaches used in applied research. The primary focus is on methods that utilise the actual values of observations, rather than their ranks. We concentrate on such desirable properties of covariograms as bias, positive-definiteness and behaviour at large distances. The paper discusses several theoretical properties and demonstrates some surprising drawbacks of well-known estimators. Numerical studies provide a comparison of representatives from different methods using various metrics. The results provide important insight and guidance for practitioners who use estimated covariograms in various applications, including kriging, monitoring network optimisation, cross-validation, and other related tasks.
}
\Plainkeywords{nonparametric estimation, covariance, semivariogram, covariogram, auto-correlation, positive-definiteness, spatial, R} %% without formatting
\Pages{1--xx}

\Address{
  Adam Bilchouris\\
  Department of Mathematical and Physical Sciences\\
  La Trobe University\\
  Melbourne, 3086, Victoria, Australia \\
  E-mail: \email{A.Bilchouris@latrobe.edu.au}\\
  URL: \url{https://scholars.latrobe.edu.au/abilchouris} \\
  
  Andriy Olenko\\
  Department of Mathematical and Physical Sciences\\
  La Trobe University\\
  Melbourne, 3086, Victoria, Australia \\
  E-mail: \email{A.Olenko@latrobe.edu.au}\\
  URL: \url{https://sites.google.com/site/olenkoandriy/} \\
}
\usepackage[T1,T2A]{fontenc}
\usepackage[utf8]{inputenc}
\usepackage[russian, british]{babel}
\DeclareFontFamilySubstitution{T2A}{lmr}{cmr}
\usepackage{babelbib}

\usepackage{amsmath}
\usepackage{amssymb}
\usepackage{amsthm}
\usepackage{amscd}
\usepackage{verbatim}
\usepackage{graphicx}
\usepackage{color}
\usepackage[mathscr]{eucal}
\usepackage{float}
\usepackage{bm}
\usepackage{import}
\usepackage{pgfplots}
\usepackage{pstricks}
\usepackage{pst-func}
\usepackage{tikz}
\usepackage{tkz-euclide}
\usetikzlibrary{external}
\usepackage{algorithm2e}
\usepackage{mathtools}
\usepackage{mathabx}
\usepackage{subcaption}

\usepackage{booktabs}
\usepackage{subcaption}
\usepackage{diagbox}

\theoremstyle{plain} %'emphasize' the following theorems
\newtheorem{theorem}{Theorem}[section]

\theoremstyle{definition} %'roman' the following theorems
\newtheorem{definition}[theorem]{Definition}

\newtheorem{estimator}{Estimator}

\newcommand{\R}{\mathbb{R}}
\newcommand{\N}{\mathbb{N}}
\newcommand{\Z}{\mathbb{Z}}

\newcommand{\CC}{\mathbb{C}}

\newcommand{\lrb}[1]{\left( {#1} \right)}
\newcommand{\lrsq}[1]{\left[ {#1} \right]}
\newcommand{\norm}[1]{\left\lVert#1\right\rVert}
\DeclarePairedDelimiter\abs{\lvert}{\rvert}
\newcommand*\mean[1]{\overline{#1}}
\newcommand\inv[1]{#1\raisebox{1.15ex}{$\scriptstyle-\!1$}}

\newcommand*{\defref}[2][]{%
	\hyperref[{def:#2}]{%
		Definition~\ref*{def:#2}%
		\ifx\\#1\\%
		\else
		\,#1%
		\fi
	}%
}

\newcommand{\iu}{{i\mkern1mu}}

\DeclareMathOperator*{\argmin}{arg\,min}
\allowdisplaybreaks

\pgfplotsset{compat=1.18}

\begin{document}

\section{Introduction}

Estimating covariograms and semivariograms of stochastic processes and random fields is an important problem in time series and spatial statistics analysis. Most applied statisticians rely on the classical estimator, as evidenced by its implementation in most R  and Python statistical packages used for the analysis of time series and spatial data, see Appendix~\ref{appRpackages}.  The standard estimator has seen routine use in statistical work due to its intuitiveness \citep[p.~174]{jenkins1969}. In fact, the correlation of a times series has been studied as early as 1905, see \cite{Cave1905}. Issues with the standard estimator were also known for some time, but its use persisted, see for instance \citet[p.~185]{jenkins1969}, \citet[pp.~71-77]{yaglom1981}, and \citet[pp.~239-245]{Yaglom1987_1}. The study of variograms is often contributed to \citet{Matheron1962}, however, \citet[p.~58]{Cressie1993} refers to papers as early as 1941 discussing the same idea, albeit under different names. Matheron's variograms were originally used in geostatistics but have now seen use in spatial statistics as a whole, see the references in the following sections. The literature now presents several alternative approaches that are less known, but are often more robust and accurate estimates with desirable properties, as will be demonstrated in the paper. To the best of our knowledge, there are no comprehensive overview publications thoroughly discussing and comparing these methods. Thus, the initial aim of this paper was to collect and compare all known approaches, however, the literature search revealed that it was an overly ambitious task for a single paper. Consequently, we restricted the scope of this manuscript to nonparametric estimators based on actual observations. Other significant areas, such as methods based on ranks and parametric approaches, are only briefly mentioned.

The paper discusses several different estimators of the covariance function known in the literature or variations upon well-known estimators. We tried to avoid the discussion of various technical modifications but rather presented nonequivalent methods or scenarios. 
Throughout the paper, we will use the terms \textit{covariance}, \textit{autocovariance}, and \textit{covariogram} interchangeably.

First, when presenting these methods, the main emphasis was on several important properties of these estimators, including bias and positive-definiteness. The primary concern was homogeneous and isotropic random fields and stationary time series, see \citet{Ma_C}, characterised by covariance functions dependent only on distances. Nonetheless, most of the materials presented are applied or can be extended to the estimation of homogeneous and directional dependencies. This paper primarily discusses issues important for applications. For proofs and theoretical results concerning various models of positive-definite functions and theoretical inference, we recommend, for example, \citet{Yadrenko}, \citet{Ivanov1989}, and \citet{Porcu}. 

Secondly, simulation studies were conducted for the mentioned methods, wherein their performance was compared using several covariance models with different properties. This included Gaussian, Bessel, and Cauchy covariance models, which exhibit short-, cyclic, and long-range dependencies. When it was possible, in this analysis, we utilised some available R packages. However, as most non-standard methods were only available in publications without code, we developed the corresponding R code to illustrate them and compare their performance. The comparison methods included metrics that compared distances between the theoretical covariogram, from which random processes were simulated, with the aforementioned nonparametric estimators. Also, a cross-validation approach based on kriging that utilised the estimated functions was used.

Most theoretical papers primarily focus on the asymptotic properties of estimators. It is well-known and rigorously proved in the literature that, within a fixed distance range, many estimators exhibit desirable asymptotic properties. However, the analysis presented in this paper reveals that if the distance range in the estimated covariogram is more than 10-20\% of the total available distance, the performance of some estimators, including classical ones, yields unreliable results. This supports the idea presented by \citet[p.~237]{Yaglom1987_1}. Also, it was demonstrated that accurate estimation of covariograms of long-range dependent data necessitates much longer distances than those typically used for estimation intervals. Consequently, most approaches that reliably handle short-range dependent data are inadequate for long-range dependent scenarios. Furthermore, this study demonstrates that for cases involving cyclic behaviours of the covariogram, certain foundational principles of well-established approaches are invalid, potentially leading to unreliable results.

The paper is structured as follows. Section~\ref{sec2} presents two key functions for estimating dependencies in the literature, covariograms and semivariograms. Section~\ref{sec:standard_method} overviews some principles of constructing and properties of the most frequently used classical estimator. Sections~\ref{sec4}   through \ref{sec9} introduce and analyse the main nonparametric estimation approaches examined within this study. Section~\ref{sec10} provides a brief overview of alternative approaches for estimating temporal and spatial dependencies in the literature. Section~\ref{sec11} presents numerical and simulation studies that compare the considered methods. Finally, the concluding section offers conclusions and outlines some problems for future research. The references include key publications that can serve as starting resources for exploring the considered methods. Some technical proofs and complimentary results are included in the Appendices.

Simulations, numerical computations and plotting in the paper were performed using the software R (version 4.1.0). The code is freely available in the folder
`Research materials' from the website \url{https://sites.google.com/site/olenkoandriy/}.

%% include your article here, just as usual
\section{Semivariogram or covariogram?}\label{sec2}
In spatial statistics, there are two similar functions used in the analysis of the dependency structure of random fields, the semivariogram, and the covariance function.

The {\it semivariogram} characterises the second-moment structure of the increments of the random field, i.e. \[\gamma(\bm{h}):= \frac{1}{2} \text{Var}(Z(\bm{s}) - Z(\bm{s} + \bm{h})),\quad \bm{s}, \bm{h}\in \mathbb{R}^d,\] whereas
the {\it covariance function} is a measure of spatial covariance, \[C(\bm{h}):=E\lrb{ (Z(\bm{s}) - \mu) ( Z(\bm{s} + \bm{h}) - \mu) }, \quad \mbox{where}\ \mu:=E(Z(\bm{s})).\]

There are several reasons why someone may want to use the semivariogram over the covariance function when performing estimation and analysis. For example,
the semivariogram estimator does not require the mean in its estimation. % \cite[pp.~136-137]{Schabenberger2004}. 
The mean, if not known, is estimated from the data, and if the data is spatially correlated, a bias may be present in the standard estimate of the mean, and thus the
estimated covariance function \citep[p.~103]{Hristopulos2020}.
Under the assumption of weak stationarity, a Gaussian random field can be defined entirely by its first and second moments, i.e. its mean and covariance \citep[p.~17]{Chils2012}.
Under weak stationarity, the semivariogram and covariance functions are related through 
\begin{equation}\label{g-c}
\gamma(\bm{h}) = C(\bm{0}) - C(\bm{h}).
\end{equation}
However, under intrinsic stationarity, this does not hold, unless the semivariogram is bounded by a finite value \citep[p.~31]{Chils2012}.

The following sections employ both functions, where it is more convenient or depending on how they were introduced in the literature, with some bias towards the covariance function to discuss positive-definiteness.% properties.

\section{Standard estimators}
\label{sec:standard_method}
The estimation of the spatial covariance function generalises the construction for the time series autocovariance function, aiming to utilise all points $\bm{s}, \bm{t} \in \mathbb{R}^{d}$ with the separation vector $\bm{h} \in \mathbb{R}^{d}$, i.e. $\bm{s} - \bm{t} = \bm{h}$.
\begin{estimator} ({\it standard or classical estimated covariance function}) \citep*[pp.69-70]{Cressie1993} \\ % Empirical Spatial Covariance Function \parencite[pp.69-70]{Cressie1993} \\
	\label{def:empSpatialCorrelation}
	The empirical spatial covariance function is given by:
	\begin{align*}
		\widehat{C}(\bm{h}): = \frac{ 1 }{ \abs{N(\bm{h})} } \sum_{N(\bm{h})} \lrb{  X(\bm{t}_{i}) - \mean{X} } \lrb{ X(\bm{t}_{j}) - \mean{X} }, \quad  \bm{t}_{i} \in \R^{d},\ i=1,...,n,
	\end{align*}
	where $N(\bm{h})$ denotes all pairs of points $(\bm{t}_{i}, \bm{t}_{j})$ whose difference is equal to the vector $\bm{h}$,
	$$
	N(\bm{h}): = \{ (\bm{t}_{i}, \bm{t}_{j}) : \bm{t}_{i} - \bm{t}_{j} = \bm{h}, i, j = 1, \dots, n \} \, ,
	$$
	and $\mean{X}$ denote the sample mean $	\mean{X}: = n^{-1} \sum_{i=1}^{n} X(\bm{t}_{i}) \, .
	$
\end{estimator}

This estimate $\widehat{C}(\bm{h})$ is also called the {\it covariogram} and is a method of moments estimator for the covariance function.

\begin{estimator} ({\it standard semivariogram estimator}) \\
    The corresponding estimator of the semivariogram proposed by \citet{Matheron1962} is
    \label{eq:variog_est}
	\begin{align*}
	\widehat{\gamma}(\bm{h}): = \frac{ 1 }{ 2\abs{N(\bm{h})} } \sum_{N(\bm{h})} \lrb{  X(\bm{t}_{i}) - X(\bm{t}_{j}) }^{2}.
	\end{align*}
\end{estimator}

Often there might not be enough pairs of points whose difference is $\bm{h}$, so one may consider adding some neighbourhood around $\bm{h}$,
controlled by $\bm{\varepsilon}$, which
can increase the number of pairs of points for the lag $\bm{h}$ \citep[p.~153]{Schabenberger2004}. A method to select this neighbourhood, for the isotropic case, will be presented in the next subsection.
Computationally, this estimator is not ideal as one must determine which pairs of points have a separation vector equal to $\bm{h}$, 
which requires searching all possible pairs of points for each $\bm{h}$.

For an isotropic covariance function, this estimator is modified by $\widehat{C}(\tau)$ that uses points $\bm{t}_{i}$ and $\bm{t}_{j},$ whose separation vector has the length $\tau$.
This is equivalent to finding all points which fall on a circle of radius $\tau$ around each point $\bm{t}_{i}$.
%For example, this method is used in the RandomFields package %, although the documentation states only cross-covariance function between two random fields
%\citep[\textbf{\texttt{RFcov}} documentation]{RFpackage}.

Appendix~\ref{appPseudocode} provides a pseudocode for computing the empirical isotropic covariance function through the above method. For simplicity, it is assumed that the mean is zero, and the covariance function is estimated for some lag distance $\tau$. For each distance $\tau$, this algorithm has a time complexity of $\mathcal{O}(N^{2}).$

For a sufficiently sampled random field on some bounded domain there are usually a high number of points which can be
considered when computing the empirical covariance function at short distances, however, if the distance is too small, there may be no pairs of points. For larger distances, the number of points decreases, as points near the edges of the domain can contribute only in certain directions.  As a result, one must carefully consider the distances they wish to use when estimating the covariance function.

% NEW
When the domain size of a random field remains the same but the number of points increases, which is called \textit{infill sampling}, the statistical properties of Estimator~\ref{eq:variog_est} change. For example, it is not consistent, see \citet{Lahiri1996}. \citet*{Lahiri2002} and \citet*{Lee2002} consider estimators using spatial subsampling to fit a variogram model. The proposed regions can cover a wide variety of shapes such as spheres, ellipsoids, and star-shaped sets \citep*[p.~839]{Lee2002}. Furthermore, \citet*[p.~75]{Lahiri2002} state when infill sampling is present in a region, it leads to strong dependence amongst the observations, similar to that of long-range dependence. This can cause issues when performing estimation, such as slower convergence to regression parameters.
% end NEW

\begin{estimator} ({\it directional estimated covariance function}) 
A variation of the standard estimator exists, called the directional estimated covariance function, which considers a specific direction only. In this case, to get enough points along the selected azimuthal angle, a conic region (called a search cone) is used \citep[p.~310]{Ma2019}. The width of the cone depends on angle tolerance and bandwidth.
\end{estimator}
In the case of an isotropic random field, the directional estimated covariance functions should be similar regardless of the angles.
For anisotropy, there should be a difference between some directions, assuming the anisotropy is strong enough.

\subsection{On some properties of standard estimators}\label{sec_prop}
Now, let us discuss some desirable properties of covariance function estimators.

A covariance function must be {\it positive-definite,} i.e.
\begin{definition}
    A function $f : \mathbb{R}^{d}\times \mathbb{R}^{d} \rightarrow \CC$ is called positive-definite if for any $n \in \N$,  $\bm{t}_{1} , \dots , \bm{t}_{n} \in \mathbb{R}^{d}$ and $a_{1} , \dots , a_{n} \in \CC$, it holds
    $$
    \sum_{i=1}^{n} \sum_{j=1}^{n} c_{i} \mean{c_{j}} f(\bm{t}_{i}, \bm{t}_{j}) \geq 0 .
    $$
\end{definition}
For a weakly stationary random field it is satisfied for $f(\bm{t}_{i}, \bm{t}_{j})=C(\bm{t}_{i}, \bm{t}_{j})=C(\bm{t}_{i} - \bm{t}_{j}).$

If an estimator is not positive-definite, then the result is not a valid covariance function, just a function that in some sense is close to a covariance function. In many cases, such estimates can be problematic, for example, kriging requires a valid covariance function~\citep{Cressie1993};  the spectral density estimates based on non-positive-definite functions may take negative values, among other potential issues.

The following properties of the classical estimator illustrate several potential issues that require careful consideration when employing it.

\begin{itemize}
	\item[(i)] 	The standard estimator does not always produce a positive-definite result, when the number of pairs of points $\abs{ N(\tau) }$ varies as $\tau$ varies, which can be shown by a simple example, see Appendix~\ref{appNonpositive}.

	\item[(ii)] In fact, the classical estimator, is always positive-definite if $\abs{ N(\tau) }$ is the same constant for all $\tau,$ see Appendix~\ref{appPositive}.
 A well-known variant of this estimator in the one-dimensional case, which is biased and positive-definite, is
    $$
        \widehat{C}^{**}(h): = \frac{1}{N} \sum_{t=1}^{N - h} (X(t) - \mean{X}) (X(t+h) - \mean{X}),
    $$
    and the associated autocorrelation estimator is
    \begin{equation} \label{eq:autocorr_est}
        \widehat{\rho}(h): = \frac{\sum_{t=1}^{N - h} (X(t) - \mean{X}) (X(t+h) - \mean{X})}{\sum_{t=1}^{N} (X(t) - \mean{X})^{2}} .
    \end{equation}

 \item[(iii)] Similar to other statistics, {\it unbiasedness and consistency} are typically desirable properties of covariance function estimators. There is a substantial theoretical literature on this topic. However, as will be demonstrated, despite the presence of pointwise convergence of the estimators for each fixed lag, the classical and many other estimators do not exhibit uniform convergence and may display undesirable behaviour in instances involving finite spatial domains or large distances.
 
 \citet{Hassani2009} and \citet{Hassani2012} provide a surprising result about the one-dimensional autocorrelation estimator, \eqref{eq:autocorr_est}, when summing over all lags $h \geq 1$, when $N \geq 2$,
    $$
    \sum_{h=1}^{N-1} \widehat{\rho}(h) = -\frac{1}{2},
    $$
    which holds regardless of observed values.
    This has implications when considering long-memory, as the sum of the estimates converges to $-1/2$, whilst it needs to be divergent in the long-memory case.

 Appendix~\ref{appSummability} proves that a similar property holds for the spatial case too. Namely,
  $$
    \sum_{\bm{h} \in H} \widehat{\rho}(\bm{h}) = -1 \quad \text{and} \quad \sum_{\bm{h} \in H} \frac{\abs{N(\bm{h})}}{N} \widetilde{\rho}(\bm{h}) = -1,
    $$
    where $H := \{\bm{h}: \bm{h} = \bm{t}_{i} - \bm{t}_{j},\ i, j = 1, \dots N,\ i \neq j \}$ is the set of all separation vectors amongst different observation locations, and $\widehat{\rho}(\bm{h})$ and $\widetilde{\rho}(\bm{h})$ denote the empirical correlations that correspond to the standard covariogram estimators with constant and varying $\abs{ N(\bm{h}) }.$
    
       The classical empirical covariograms will have more pronounced waves for the estimated values $\widetilde{\rho}(\bm{h})$ as the multiplicative weights are within $[0, 1]$. These waves, as will be shown in the simulation results, are artefacts in the estimation process caused by fewer samples being available to compute the empirical covariance function as the estimation lag increases. Furthermore, analogous identities can be obtained for other commonly used weighted estimators where the weights depend only on the separation lag (or distance) between the locations of samples, rather than the sampled values at the locations.

 	\item[(iv)] Furthermore, the classical estimator does not hold the aforementioned relation (\ref{g-c}) between the variogram and covariance function, see Appendix~\ref{appRelation}.   
\end{itemize}

\subsection{Selection scheme for the standard method}\label{selection}
When estimating the covariance function, the data one is dealing with, whether it is simulated, or real, will only be sampled finitely.
As a result, the empirical covariance function cannot be computed at all distances, except when there is a sufficient number of points
for each distance (such as in the case of functional data).
    
\begin{estimator} ({\it bin-type estimators})
	There exist several variations, even within the standard method, which specify the sampled values to be used for each particular distance $\tau$ (or the vector $\bm{h}$) and choose the extrapolation technique between distinct $\tau_i$ (or $\bm{h}_i$).
\end{estimator}

\citet{krajewski1988} provide two schemes to select the distances considered, and the error region, $\Delta,$ when estimating an isotropic
covariance function. Adding the error region is common in practice so a sufficient number of points are included for each distance $\tau$.  We will only consider selection scheme 1 in \citet{krajewski1988}, as it is simpler when working with a square lattice. The error region is constructed around a circle of radius $\tau$, resulting in a region bounded by two circles with radii $\tau + \Delta$ and $\tau - \Delta$. 
The scheme is specified as follows 
\begin{enumerate}
	\item $\tau_{k+1} - \tau_{k} = $ constant, $k=1, \dots,K-1$.
	\item $\tau_{1}: = \displaystyle \frac{1}{n} \sum_{i=1}^{n} \min\lrsq{ d_{ij} : j=1, \dots, n : j \neq i }$,
	where $d_{ij}$ is the distance from point $\bm{t}_{i}$ to $\bm{t}_{j}$.
	\item $\Delta := \tau_{1} / 2$.
	\item $K$ is fixed (usually $K$ is between 10 and 15).
\end{enumerate}
This does not allow any control for the number of pairs of points $n_{k}$ for some $\tau_{k}$. If there are insufficient samples used when computing the empirical covariance function, even in the one-dimensional case, there will be waves present, see \citet[p.~239]{Yaglom1987_1} and \citet[p.~74]{yaglom1981}.
In \citet*[p.~194]{Journel1978}, they recommend the number of pairs for each lag to be at least 30, and the maximum lag to be less than half the maximum separation lag, which were supported by the simulation studies in \citet{Cressie1988}. The choice of the maximum separation lag will determine how prominent the waves will be.

\subsection{Estimators on the ball}

It is difficult to mention numerous generalisations of the classical estimator in the literature. We only provide an example, although the case of sampled functional data is not considered here, of the following estimator by \cite{Dyhovichnyj1983} for the cases of functional data with observation windows that include a ball. It is the continuous equivalent of Estimator~\ref{sec:standard_method}. 
\begin{estimator} ({\it classical estimated covariance function for functional data})
	An estimator of a covariance function of a zero-mean isotropic random field on an $n$-ball of radius $R$ centred at the origin, $V_{R}(0),$ is given by
	$$
	\widehat{C}(r): = \frac{1}{U_{n}(R - r)} \int_{V_{R - r}(0)} X(\bm{s}) \left( \frac{1}{\omega_{n}(r)} \int_{S_{r}(\bm{s})} X(\bm{t}) m_{n}^{(r)} (\text{d}\bm{t})\right) \text{d}\bm{s},
	$$    
	where $r$ is the distance lag, $S_{r}(\bm{s})$ is a sphere of radius $r$ centred at $\bm{s}$, $U_{n}$ and $\omega_{n}$ are the volume of the ball and surface area of the sphere,
	respectively, and $m_{n}^{(r)}$ is a Lebesgue measure on the sphere.
\end{estimator}

\section{Variogram-based estimators}\label{sec4} 
Although we are mainly concerned with covariance function estimation, it is important to note that in the cases when it is appropriate (see the discussion in (iv) of Section~\ref{sec:standard_method}), one can employ the following approach. 
\begin{estimator} ({\it variogram-based estimated covariance function})
	\[\widehat C(\bm{h}) = \widehat C(\bm{0}) - \widehat\gamma(\bm{h})=\widehat{\text{Var}}(X) - \widehat\gamma(\bm{h})\] 
	where $\widehat\gamma(\bm{h})$ is a variogram estimator known in the literature.
\end{estimator}  For example, one can use the popular estimator proposed by \citet*{Cressie1980} 
$$
2\widehat{\gamma}(\bm{h}): = \lrb{ \frac{1}{N(\bm{h})} \sum_{i=1}^{N(\bm{h})} \abs{ X(\bm{t}_{i + \bm{h}}) - X(\bm{t}_{i}) }^{1/2} }^{4} {\bigg/} (0.457 + 0.494 / N(\bm{h})) ,
$$
or its robust generalisation in \citet[p.~214]{Genton1998}.

\section{Kernel regression estimators}
Below is an estimator similar to that of Nadaraya-Watson kernel regression, proposed by \citet{hall1994_2} and \citet{hall1994}.
\begin{estimator} \label{eq:hall_1}({\it kernel-based estimated covariance function})
\begin{align*}
	\widehat{C}_{H}(\bm{t}) := \frac{ \displaystyle \sum_{i} \sum_{j}  \widecheck{X}_{ij} K\lrb{ (\bm{t} - (\bm{t}_{i} - \bm{t}_{j})) / b } }{ \displaystyle \sum_{i} \sum_{j} K\lrb{ (\bm{t} - (\bm{t}_{i} - \bm{t}_{j})) / b } },
\end{align*} where $\widecheck{X}_{ij}: = (X(\bm{t}_{i}) - \mean{X} )( X(\bm{t}_{j}) - \mean{X}).$ 
\end{estimator}
One has a choice for the kernel function $K(\cdot)$ used, however, it must have the properties of symmetric probability density, resulting in $\widehat{C}_{H}(-\bm{t}) = \widehat{C}_{H}(\bm{t})$ \citep*[p.~403]{hall1994}.

A method in the paper by \citet{hall1994} was proposed to ensure the positive-definiteness of the estimator $\widehat{C}_{H}$, by manipulating its Fourier transform, $\mathcal{F}( \widehat{C}_{H}(\bm{t}) ),$ which forces negative values of this Fourier transform to zero before performing a Fourier inversion. 

The approach can be utilised in more general settings, where an estimate of the covariance function is a non-positive-definite function. Namely,
\begin{estimator}({\it kernel-based positive-definite estimated covariance function}) ${}$
\begin{enumerate}
\item Compute $\widehat{C}_{H}(\bm{t})$.
	\item Compute $\mathcal{F} ( \widehat{C}_{H}(\bm{t}) )$.
	\item Set $\widehat{\mathcal{F}} ( \widehat{C}_{H}(\bm{t}) ) = \mathcal{F}( \widehat{C}_{H}(\bm{t}) )$ if $\mathcal{F} ( \widehat{C}_{H}(\bm{t}) ) > 0$, and $\widehat{\mathcal{F}} ( \widehat{C}_{H}(\bm{t}) ) = 0$ otherwise, for every frequency $\bm{\theta}$.
	\item Compute the inverse Fourier transform of $\widehat{\mathcal{F}}( \widehat{C}_{H}(\bm{t}) )$ and use it as a new estimator  $\widetilde{C}(\bm{t}).$
\end{enumerate}
\end{estimator}

A related approach was provided in the paper by \citet*[p.~2118]{hall1994_2}, which considered forcing the estimated covariance function down to zero linearly starting at the point $T_{1} > 0$ and ending at $T_{2} > T_{1}.$ For example, for Estimator~\ref{eq:hall_1}, $\widehat{C}_{H}(t),$ in the one-dimensional case, it gives:
\begin{estimator} ${}$ ({\it kernel-based truncated estimated covariance function})
\begin{enumerate}\label{eq:hall}
	\item Compute the empirical covariance function as
	$$
	\widehat{C}_{1}(t): =
	\begin{cases}
		\widehat{C}_{H}(t), & 0 \leq t \leq T_{1} \\
		\widehat{C}_{H}(T_{1}) (T_{2} - t) (T_{2} - T_{1})^{-1}, &  T_{1} < t \leq T_{2} \\
		0 & t > T_{2}.
	\end{cases},
	$$
	\item Perform a Fourier transform (which is a Fourier-cosine transform due to the real symmetric nature of $\widehat{C}_{1}$)
	$$
	\mathcal{F}^{c} \lrb{ \widehat{C}_{1}(t) } = 2 \int_{0}^{\infty} \cos(\theta t) \widehat{C}_{1}(t) \, \text{d} t .
	$$
	\item Determine $\widehat{\theta} = \inf \left\{ \theta > 0 : \mathcal{F}^{c} \lrb{ \widehat{C}_{1}(t) } < 0 \right\}$. Then, set all values of
	$\mathcal{F}^{c} \lrb{ \widehat{C}_{1}(t) },$ when $\theta > \widehat{\theta},$ to zero.
	\item Perform an inversion to obtain \[\widetilde{C}(t): = \int_{-\widehat{\theta}}^{\widehat{\theta}} (2\pi)^{-1} \mathcal{F}^{c} \lrb{ \widehat{C}_{1}(t) } \cos(\theta t) \, \text{d}\theta.\]
\end{enumerate}
 
\end{estimator}

This method and its resulting estimator will be referred to as {\it Hall's estimator.}

One issue with bringing the tail of the estimator down to zero is it can no longer be used for long-range dependent random fields (or long-memory processes), as the absolute values of their covariance functions are non-integrable. This is the same issue as Estimator~\ref{eq:yaglom_3_2}, which will be introduced later, although this can help to remove the waves mentioned earlier.
Also, the last method discussed in this section, making the estimated function positive-definite, may fail if $\widehat{\theta}$ is very close to zero.

% NEW
The kernel-based estimated covariance function has been adapted in the literature for semivariogram estimation. For example, \cite{GarciaSoidan2004} considers a variant to estimate an isotropic semivariogram, \cite{GarciaSoidan2012} considers it for an indicator variogram estimator, \cite{cuevas2013study} construct a cross-variogram estimator, and \cite{yu2007kernel} construct a variable nearest neighbour estimator for the semivariogram.
% end NEW

% Add Genton's.
\section[Qn estimators]{$Q_{n}$ estimators} \label{Qn}

The papers by \citet{Genton1998} and \citet{Genton2000} provide a robust estimator for the variogram and covariance function based on the $Q_{n}$ estimator of \citet*{Rousseeuw1993}.
$Q_{n}$ is an estimator of scale, and has a breakdown point of 50\%, meaning up to 50\% of observations can be outliers before an incorrect (arbitrarily large) result is given \citep[p.~1277]{Rousseeuw1993}. Unfortunately, the covariance version of this estimator only has a breakdown point of 25\%, which is on par with the interquartile range (\citeauthor{Rousseeuw1992} \citeyear{Rousseeuw1992}, p.~78; \citeauthor{Genton2000} \citeyear{Genton2000}; p.~655).
$Q_{n}$ is defined as follows:	
$$
Q_{n}(\bm{X}) := c \{ \abs{X_{i} - X_{j}}; i < j \}_{(m)} ,
$$
where $\bm{X}$ is a vector of observations $(X_{1}, \dots , X_{n})^{\prime}$ , $c$ is a constant for consistency, and
$
m = \lrsq{ \left({{\binom{n}{2}} + 2}\right)/{4} } + 1 ,
$ where $\lrsq{ \cdot }$ denotes the integer part.
Thus, it computes the $m^{\text{th}}$-order statistic of the absolute differences of the observations.  For large $n$, this is approximately the first quartile, and in the case of a Gaussian distribution, the consistency constant can be selected as $c=2.2191$ \citep* [p.~655]{Genton2000}. For $Q_{n}$, the outliers refer to the absolute pairwise differences, not the observed values in $\bm{X}$. 

In the variogram case, \citet*[pp.~214-216]{Genton1998} considers constructing a random field of differences at the spatial lag $\bm{h}$, $V(\bm{h}) = X(\bm{t} + \bm{h}) - X(\bm{t})$, which has zero mean and variance $2\gamma(\bm{h})$. From this, a sample $\{ V_{1}(\bm{h}) , \dots , V_{\abs{N(\bm{h})}} \}$ of $V(\bm{h})$ which corresponds to the sample $\{ X(\bm{t}_{1}) , \dots , X(\bm{t}_{N}) \}$ of $X$, gives
$$
Q_{\abs{N(\bm{h})}} := 2.2191 \{ \abs{ V_{i}(\bm{h}) - V_{j}(\bm{h}) } \}_{(m_{S})} ,
$$
where $m_{S} = \binom{ [\abs{N(\bm{h})} / 2] + 1}{2}$, and finally, $2\widehat{\gamma}(\bm{h}) = ( Q_{\abs{N(\bm{h})}} )^{2}$ \citep*[pp.~214-216]{Genton1998}.

The estimator about to be introduced, and the estimator to follow, are motivated by the following identity, which assumes finite second-moment (i.e. $X, Y \in L^{2}$) \citep*[p.~202]{Huber1981}:
$$
\text{Cov}(X, Y) = \frac{1}{4ab} \lrb{ \text{Var}(aX + bY) - \text{Var}(aX - bY) } ,
$$
where $a, b$ are arbitrary constants, although they should be chosen on a similar scale to avoid strange results.
In the following two estimators, as we are considering processes with unit variance, we have $a = b = 1.$

\begin{estimator} ({\it quantile-based estimated covariance function}) \citep*[p.~665]{Genton2000} \\
	\label{eq:genton_est}
	The $Q_{n}$ estimator for the covariance function is given by
	\begin{align*}
		\widehat{C}_{Q}(h, \bm{X}): = \frac{1}{4} \lrb{ Q^{2}_{n - h} (\bm{X}_{1:(n-h)} + \bm{X}_{(h+1):n}) - Q^{2}_{n - h}(\bm{X}_{1:(n-h)} - \bm{X}_{(h+1):n}) },
	\end{align*}
	where the vector $\bm{X}_{1:(n-h)}$ consists of first $n-h$ observations of the vector $\bm{X}$ and the vector $\bm{X}_{(h+1):n}$  includes last $n-h$ observations.    
\end{estimator}

For this estimator, the autocorrelation function cannot be obtained from the autocovariance in the usual way, $\rho(h) = C(h)/C(0)$, as this is not bounded between $-1$ and $1$. Instead, \citet[p.~666]{Genton2000} provide the following robust estimate.
\begin{estimator} ({\it quantile-based estimated correlation function})
$$
\widehat{\rho}_{Q}(h, \bm{X}) := \frac{ Q^{2}_{n - h} (\bm{X}_{1:(n-h)} + \bm{X}_{(h+1):n}) - Q^{2}_{n - h}(\bm{X}_{1:(n-h)} - \bm{X}_{(h+1):n}) }{ Q^{2}_{n - h} (\bm{X}_{1:(n-h)} + \bm{X}_{(h+1):n}) + Q^{2}_{n - h}(\bm{X}_{1:(n-h)} - \bm{X}_{(h+1):n}) } .
$$
\end{estimator} 
Unlike other estimators, this estimator is location-free, as it does not depend upon any knowledge of the location of the points.
For example, the standard estimator, Estimator~\ref{def:empSpatialCorrelation}, depends on the locations between points, whereas this one considers pairwise differences.
This estimator has a breakdown point of 25\%, which is the highest possible value when considering autocovariance \citep*[p.~665]{Genton2000}.

\subsection[Pn estimator]{$P_{n}$ estimator}
Another robust estimator was proposed by \citet{TarrSlides}, based on the $P_{n}$ estimator of pairwise means in \citet{Tarr2012}, and is related to U-statistics.
\begin{estimator} ({\it pairwise mean-based estimated covariance function})
$$
\widehat{C}_{P}(h): = \frac{1}{4} \lrb{ P^{2}_{n - h} (X_{1:(n-h)} + X_{(h+1):n}) - P_{n - h}^{2} (X_{1:(n-h)} - X_{(h+1):n}) } ,
$$
where $P_{n} = c \lrb{ M_{n}^{-1}(0.75) - M_{n}^{-1}(0.25) }$, with $c \approx 1.048$ being a correction factor ensuring $P_{n}$ is consistent for the standard
deviation when the observations are Gaussian. $M_{n}(\cdot)$ denotes the empirical distribution of pairwise means,
$$
M_{n}(t): = \frac{2}{n(n-1)} \sum_{i < j} \bm{1}\left(\frac{X_{i}+ X_{j}}{2} \leq t\right),\quad  t \in \R,
$$
and $\bm{1}(\cdot)$ denotes the indicator function.
\end{estimator}

$P_{n}$ has a breakdown point of 13.4\%, whilst $Q_{n}$ has a breakdown point of 50\%, although for Gaussian asymptotic efficiency, $P_{n}$'s is $0.86$ whereas $Q_{n}$'s is $0.82$
(\citeauthor*{Tarr2012} \citeyear{Tarr2012}, p.~193; \citeauthor*{TarrSlides} \citeyear{TarrSlides}).

% Edge effect
\section{Tapered estimates}
\subsection{Edge effect}
An important factor to consider when working with stochastic processes, especially as the dimension increases, is the \textit{edge effect}.
Let $D$ be some study domain, such as a finite rectangular lattice $P_{n}: = \{ 1, \dots, n_{1} \} \times \cdots \times \{ 1, \dots, n_{d} \}$ in $d$-dimensional integer space, $\Z^{d}$. The points on the boundary of $D$ will have their
nearest neighbours lying outside of $D$, meaning observations are unavailable at those locations. As $d$ increases, so does the number of boundary points. For example, for $N=100$ points, a one-dimensional interval has two points, or 2\%, on the edge,
but for a square with side lengths $10$, there are 36 points on the boundary or 36\%. This number quickly increases
as $d$ increases \citep*[p.~478]{Cressie1993}. 

As a result, the edge effect also introduces bias, so one must be aware of its presence \citep*[p.~607]{Cressie1993}.
There are many techniques to deal with edge effects in practice, such as treating a rectangular study region as a torus \citep*{Griffith1983}.
In practice, edge effects can arise in unexpected scenarios, such as when patients receive healthcare outside of a study area \citep*{Fortney2000}.

\subsection{Tapered covariance estimators}
To deal with the edge effect several corrections of standard statistical estimators (tapered estimators) were proposed.
For example, \citet*{Guyon1982} suggested to use the following unbiased estimator
\begin{estimator}({\it tapered unbiased estimated covariance function})
	$$
	\widehat{C}_{N}(\bm{h}) := \frac{1}{ C_{N - \abs{\bm{h}}} } \sum_{\bm{t}, \bm{t} + \bm{h} \in P_{n}} X(\bm{t}) X(\bm{t} + \bm{h}) ,
	$$
where $C_{N - \abs{\bm{h}}}$ is the cardinality of the set $\{ \bm{t} : \bm{t}, \bm{t} + \bm{h} \in P_{n} \}$ and  $N=\prod_{i=1}^{d}n_{i}.$ 
\end{estimator}
However, it was shown that this estimator is not positive-definite and has some other potential issues.
One of the corrections, which properly takes the edge effect into account was proposed by \citet{Dahlhaus1987}:
\begin{estimator} (\textit{tapered estimated covariance function})
    \label{eq:tapered}
	\begin{align*}
	\widehat{C}^{a}_{N}(\bm{h}): = \lrb{ \prod_{i=1}^{d} H_{2, n_{i}}(0) }^{-1} \sum_{\bm{t}, \bm{t}+\bm{h} \in P_{n}} X(\bm{t}) X(\bm{t}+\bm{h}) \lrb{ \prod_{i=1}^{d} a( (t_{i} - 1/2)/n_{i}; \rho ) a( (t_{i} + h_{i} - 1/2)/n_{i}; \rho ) } .
	\end{align*}
		Here $a(\cdot; \cdot)$ is a taper (window) function with smoothness parameter $\rho$ and $u \in [0, 1]$,
	$$
	a(u; \rho) := 
	\begin{cases}
		w(2u/\rho) ,& 0 \leq u < \frac{1}{2}\rho, \\
		1 ,& \frac{1}{2} \rho \leq u \leq \frac{1}{2}, \\
		a(1-u; \rho) ,& \frac{1}{2} < u \leq 1,
	\end{cases} 
	$$
	where $w(\cdot)$ is a continuous increasing function with $w(0)=0$ and $w(1)=1.$
	Additionally,
	$$
	H_{2, n}(0) := \sum_{s=1}^{n} a ( (s - 1/2)/n; \rho )^{2} .
	%H_{j, n}(\alpha) = \sum_{s=1}^{n} h\{ (s - 1/2)/n) \}^{j} \exp(-i\alpha s) .
	$$
\end{estimator} 

A possible choice for the function $w(\cdot)$, as suggested by \citet*[p.~878]{Dahlhaus1987}, is the Tukey window, although one is not limited to this choice. 

This estimator is biased, although it becomes asymptotically negligible due to the use of windowing. As $H_{2, n}(0)$ is a constant, a similar idea can be applied as in the standard estimator to show the tapered estimator is positive-definite. Additionally, it 
also can be applied in non-Gaussian scenarios \citep*[p.~403]{Yao2006}.

Much like Hall's estimator, this incorporates windowing on the standard estimate of the covariance, under the assumption of a zero-mean process.
The incorporation of a taper does not inhibit this estimator's ability to consider long-range dependent processes to the same extent as Hall's estimator and Estimator~\ref{eq:yaglom_3_2}, which will be introduced soon.

\section{Estimates via linear combinations of basis functions}\label{sec:splines}\nopagebreak
An estimator for the isotropic autocovariance function can be constructed closely matching an empirical estimator with a linear combination of properly selected basis functions. 

The first approach uses the fact that an isotropic function $C(h)$  is positive-definite for all
$d\ge 1,$ if and only if $\phi(h) = C(h^{1/2})$ is completely monotone \citep[p.~86]{Cressie1993}.

Let $\{ B_{1}^{(p)} , \dots, B_{m + p + 1}^{(p)} \}$ denote $m + p + 1$ B-splines of order $p$, with $m$ equally spaced knots, with end knots $k_{0} = 0$ and $k_{m} = 1$.
Also, let $\{ b_{1} , \dots , b_{m + p + 1} \}$ denote the B-spline coefficients.

\citet*[p.~615]{Choi2013} provided the following representation for $\phi(\cdot)$ using the B-splines:
$$
\phi(x) = \sum_{j=1}^{m + p} \beta_{j} f_{j}^{(p - 1)} (x) ,
$$
with
$$
f_{j}^{(l)} (x) = \int_{0}^{1} (m + 1) t^{x} B_{j + 1}^{(l)} (t) \text{d} t ,
$$
where $\beta_{j} = b_{j+1} - b_{j} \geq 0$ for all $j = 1, \dots , m + p$.
The functions $f_{j}^{(l)}$ are completely monotone and serve as basis functions.

Therefore, the covariance estimator can be given as 
\begin{estimator} \label{eq:splines_est} ({\it linear combination of completely monotone functions})\citep*[p.~615]{Choi2013}
  \begin{align*}
    \widehat{C}^{B}(\tau) = \sum_{j=1}^{m + p} \beta_{j} f_{j}^{(p - 1)} (\tau^{2}). 
  \end{align*}    
\end{estimator}

The coefficients $\bm{\beta} = ( \beta_{1} , \dots , \beta_{m + p} )^{\prime}$ are estimated using the weighted least squares (WLS) approach,
$$
\hat{\bm{\beta}}_{\text{WLS}} = \argmin_{\beta_{j} \geq 0} \sum_{i = 1}^{L} w_{i} \lrb{ \widehat{C}(\tau_{i}) - \sum_{j=1}^{m + p} \beta_{j} f_{j}^{(p - 1)} (\tau_{i}^{2}) }^{2} ,
$$
where $\widehat{C}(\cdot)$ is an estimator of a covariance function,  $\{ \tau_{1} , \dots , \tau_{L} \}$ is a set of distance lags, and $\{ w_{1}, \dots , w_{L} \}$ is a set of weights. The standard estimator, Estimator~\ref{def:empSpatialCorrelation}, was used in \citet*[p.~617]{Choi2013}.

As mentioned earlier, some estimators suffer from oscillations, meaning the estimator for the isotropic autocovariance function can become unpredictable as the distance lag changes, particularly depending on how many points are separated by a distance $\tau$. To address this issue, either a reasonably short range of $\tau$ is selected, or the set of weights is chosen based on the number of available points when computing the empirical autocovariance function.
For example, \citet*[p.~617]{Choi2013} recommended the following weights: $w_{i} = \abs{N(\tau_{i})} / (1 - \widehat{C}(\tau_{i}))^{2}.$
Furthermore, one is not limited to the choice of the standard estimator of the autocovariance function when selecting the weights and the fitting process.

An isotropic covariance function in $\R^{n}$ has the following form, see~\citet{Yadrenko}, and \citet{Ivanov1989},
$$
C(\tau) = \int_{0}^{\infty} \Lambda_{n}(k \tau)  \, \text{d}F(k) ,
$$
where $F(\cdot)$ is a cumulative distribution function,
$$
\Lambda_{n}(k \tau) = 2^{\frac{n-2}{2}} \Gamma\lrb{ \frac{n}{2} } \frac{J_{\frac{n-2}{2}} (k\tau)}{(k\tau)^{\frac{n-2}{2}}} 
$$
and $J_{\nu}(\cdot)$ denotes the Bessel function of the first kind of the order~$\nu.$

\citet*[p.~12]{Li2023} considered an approximation of $F(\cdot)$ by a step function, with jump points at~$k_{i}$, resulting in the following nonparametric estimator of the isotropic covariance function.
\begin{estimator}({\it linear combination of isotropic basis functions})
$$
\widehat{C}(\tau): = \sum_{i = 1}^{m} w_{i} \Lambda_{n}(k_{i}\tau) ,
$$
where $\Lambda_{n}(\cdot)$ is acting as a basis function and $\{w_{i}\}$ are weights.
\end{estimator}
The locations of the jumps $\{ k_{i} \}$ are determined in an ad hoc manner, although some methods exist such as selecting them to be roots of the Bessel function of the first kind \cite[p.~13]{Li2023}.
% NEW
The above estimator is related to the semivariogram estimator in~\citet*{Shapiro1991}, although considered for the autocovariance case instead.
The method in~\citet*{Shapiro1991} has some limitations, such as the number of jumps having no upper limit, or how to nonparametrically estimate the nugget, sill, and range \citep[p.~25]{Cherry1994}.
% END NEW

\citet*{Wang2023} considered the next estimator using a different class of basis functions.
\begin{estimator}({\it linear combination of Bernstein polynomials})
 For any dimension $d \geq 1,$   a linear combination of the basis functions
    $$
A_{i, m}(\tau) = \prod_{j=i}^{m} \lrb{ 1 + \frac{\tau^{2}}{j} }^{-1},\quad 1 \leq i \leq m,\quad i, m \in \N,
$$
provides a valid empirical isotropic covariance function 
$$
\widehat{C}_{m}(\tau): = \sum_{i=1}^{m} w_{i} A_{i, m}(\tau).
$$
\end{estimator}

\section{Correction of established methods}\label{sec9}
When established or new methods for estimating covariograms lack the required properties, such as positive-definiteness, several general transformation strategies can be employed to achieve them. Two popular approaches are discussed in this section.
\subsection{Shrunken covariance}
Numerous papers estimate a covariance matrix and subsequently transform the estimate into a positive-definite matrix if it is not inherently so. As the sample covariogram can be derived from the estimated covariance matrix, and conversely, the estimated covariance matrix can be constructed from the sample covariogram, these transformation methods are also applicable to covariograms.

One of the popular transformations, shrinking, was introduced in \citet*{Devlin1975}.
\begin{estimator}({\it linear shrinking})
A $p\times p$ pseudo-correlation matrix $\bm{R}$ is shrunk towards the $p$-dimensional identity matrix $\bm{I}_{p},$ as
$$
\widetilde{\bm{R}} := \lambda \bm{R} + (1 - \lambda) \bm{I}_{p} ,
$$
where $\lambda \in [0, 1]$ is the largest value, which makes $\widetilde{\bm{R}}$ positive-definite.   
\end{estimator}

An alternative to linear shrinking is
\begin{estimator}({\it nonlinear shrinking})
Each off-diagonal element $r$ of $\bm{R}$ is replaced by $\tilde{r}$ through the following transformation procedure
$$
\tilde{r} :=
\begin{cases}
\inv{f}(f(r) + \Delta) ,& \text{if } r < -\inv{f}(\Delta) , \\
0 ,&  \text{if } \abs{r} \leq \inv{f}(\Delta), \\
\inv{f}(f(r) - \Delta) ,&  \text{if } r > \inv{f}(\Delta),
\end{cases}
$$
where $\Delta$ is a small positive constant (such as $0.05$) and $f : \R \rightarrow [-1, 1]$  is a monotone increasing continuous function. 
This process is repeated until a positive-definite matrix is given.
\end{estimator}
Two popular choices for $f$ are $\tanh(x)$ and $\frac{2}{\pi} \arctan(x)$ \citep*{Rousseeuw1993_2}.
Shrinking does not consider the entire correlation structure of the matrix, although it is easy to apply practically, as both linear and nonlinear shrinking operate on each correlation separately.

\subsection{Kernel correction}
One can use the following modification of the approach presented in \citet*[eq.~(1.76)]{yaglom1981} for the case of $d=1$ to correct the estimate $\widehat{C}(\bm{h}).$
\begin{estimator}({\it kernel-corrected estimated covariance function})
	\begin{equation*}\label{eq:yaglom_3_2}
		\widehat{C}^{(a)}(\bm{h}) = a(\bm{h}) \widehat{C}(\bm{h}) ,
	\end{equation*}
	where $a(\cdot)$ is a kernel function approaching to 0, when $\bm{h}$ increases.
\end{estimator}  
We will now extend some of the comments in the discussion in \citet*[eq.~(1.76)]{yaglom1981} which are also relevant to the spatial case.

As mentioned earlier, covariance function estimators suffer from waves as the number of samples used to estimate the covariance function decreases. Estimator~\ref{eq:yaglom_3_2} introduces a kernel function to remove such waves, by gradually bringing down the estimator to zero. A common choice of $a(\bm{h})$ is a function that vanishes after a certain distance, for example, approximately~$0.1$ of the observation period in the case of time series.
When working under the assumption of short-range dependence, such choices are acceptable,  but in the case of long-range dependence, this may not be an appropriate estimator.
Applying the vanished estimator to a long-range dependent process results in a short-short-range dependent covariance function, which does not accurately reflect the process' dependencies. A possible strategy is to increase the support (range) of  $a(\bm{h})$ with the observation area or the number of observations.

One can choose $a(\bm{h})$ as a positive-definite kernel to keep the estimator positive-definite, given $\widehat{C}(\bm{h})$ is a positive-definite estimator. However, it restricts the class of eligible functions one can use. \citet{Genton2002} provides
a list of some valid kernels, see Table~\ref{tab:kernels}, where $\tau=||\bm{h}||$ and $\theta > 0$ in all cases. These kernels are not only positive-definite but also isotropic, meaning they can also be applied to the estimators of isotropic covariance functions.

\begin{table}[!htbp]
	\begin{tabular}{| c | c | c |} \hline
		Name & Equation & Validity \\ \hline
		Circular & $a(\tau; \theta) = 
		\begin{cases}
			\frac{2}{\pi} \arccos ( \tau / \theta ) - \frac{2}{\pi} \frac{\tau}{\theta} \sqrt{ 1 - \lrb{ \frac{\tau}{\theta} }^{2} }   , \text{ for } \tau < \theta \\
			0 , \text{ otherwise}
		\end{cases}$ & $\R^{2}$ \\ \hline
		Spherical & $a(\tau; \theta) = 
		\begin{cases}
			1 - \frac{3}{2} \frac{\tau}{\theta} + \frac{1}{2} \lrb{ \frac{\theta}{\tau} }^{3}  , \text{ for } \tau < \theta \\
			0 , \text{ otherwise}
		\end{cases}$ & $\R^{3}$ \\ \hline
		Rational Quadratic & $a(\tau; \theta) = 1 - \frac{\tau^{2}}{\tau^{2} + \theta}$ & $\R^{d}$ \\ \hline
		Exponential & $a(\tau; \theta) = \exp(-\tau / \theta)$ & $\R^{d}$ \\ \hline
		Gaussian & $a(\tau; \theta) = \exp(- \tau^{2} / \theta)$ & $\R^{d}$ \\ \hline
		Wave & $a(\tau; \theta) = \frac{\theta}{\tau} \sin(\tau / \theta)$ & $\R^{3}$ \\ \hline
	\end{tabular}
	\caption{List of isotropic positive-definite kernels}
	\label{tab:kernels}
\end{table}

\section{Other estimators}\label{sec10}\nopagebreak
In this paper, our primary focus has been on the estimation of Pearson-type correlations, which are based on the actual values of observations. We assumed that one realisation of the stationary spatial process is available and mainly considered nonparametric estimators given their significance in addressing numerous spatial issues. Even parametric estimators are frequently derived by fitting a parametric covariance function family to a nonparametric empirical covariance function.

Several alternative or closely related nonparametric approaches exist, though beyond the scope of this discussion due to space constraints and will be only briefly mentioned.

Firstly, there are methods based not on actual values but on the signs or ranks of the data. For additional information, one can refer to \citet{Durre2015} and \citet{Sang2016}, and the comprehensive citations included within.
\citet*{Sang2016} argues that Pearson-type estimates perform poorly when dealing with heavy-tailed or asymmetric distributions, as well as being impacted greatly by outliers.
There were also several modifications of the approach considered in Section~\ref{Qn}. The majority of these modifications were based on the methods utilising a robust scale measure as proposed in the highly cited publication \citet{Gnanadesikan1972}.

There also exist several approaches that used weighted, affine equivariant estimators in combination with M-estimators, see, for example, the publications \citet*{Durre2015}, \citet{Tyler1987}, and \citet{Lumley1999} and the discussions in them. It is worth mentioning that, similar to the discussion in Appendix~\ref{appSummability}, most of these models satisfy certain summability constraints that may impact their performance.

% NEW
An overview of nonparametric Bayesian approaches can be found in \citet{Kidd2022}, which has an emphasis on methods which have nice properties computationally, namely parallelisation and scalability. A nonparametric Bayesian approach was also considered in \citet{porcu2021nonparametric}, where they studied the estimation of isotropic covariance functions for random fields on the sphere.
% end NEW

Finally, there are other options that use variogram- or spectrum-based ideas, for example, \citet{Cressie2011} used spline polynomials to estimate the isotropic spectral density, and then it was employed to compute the variogram.
Such methods that produce negative-definite variogram estimates can be easily adapted to the covariance case.

\section{Simulation studies}\label{sec11}
This section will demonstrate some of the considered properties and empirically compare the estimators via realisations of Gaussian random fields with different covariance functions. For the sake of simplicity and visualisations, we only present results for two-dimensional cases, but the corresponding one-dimensional results were also obtained.

For illustration, we will consider the three isotropic covariance functions, namely 
\begin{itemize}
	\item {\it Gaussian:} $C(\tau; \sigma) = \exp(-\tau^{2} / \sigma^{2}),$ where $\sigma>0,$ %{\color{red} Please add the range parameter in the formula and refer to it later using its variable, rather than the word "range." I believe it is more appropriate to call this parameter the "scale" parameter, as spatial statistics use "range" for other purposes. You may recall this from SPA subjects. However, if "range" is the term used in the literature as you mentioned, we should call it "range" here just once and not use "range" later to avoid confusion {\color{blue} I will just add the parameter $a$ for the Gaussian case and not name it anything, as it did not get used in the other models (it was just always $1$). Renamed to $r$.}}, {\color{red} can you use sigma, $r$ was used before at least 2 times for different things. I remove sigma from the following spectral norm. We experience a shortage of letters :-)}
    \item {\it Bessel:} $C(\tau; \nu) = 2^{\nu} \tau^{-\nu} \Gamma(\nu + 1) J_{\nu}(\tau),$ where $\nu \geq (d-2)/2$ \label{eq:besselCov},
\item {\it Cauchy:} $C(\tau; \gamma) = (1 + \tau^{2})^{-\gamma}$, where $\gamma > 0.$
\end{itemize}
They are positive-definite in all dimensions $d$.%, {\color{red} given, the Euclidean distance is used for the lag} \citep[pp.~133-134]{Hristopulos2020}.

For each covariance function, a single spatial realisation will be used, as it can be deemed more realistic in many applications. To obtain reliably estimated values, these realisations have been generated either in a large observation window or a dense sampling grid. The analysis can also be extended to the methods which use multiple realisations (often modelled as independent realisations of random fields with the same statistical properties), for example, see \citet[Section~4.5]{Ivanov1989}. 

For the convenience of the readers, we reiterate the previous notations for the estimators to be discussed in this section:
\begin{itemize}
\item $C^{*}$, the standard estimator, Estimator~\ref{def:empSpatialCorrelation} with the normalising factor varying with lags, %the standard estimator, where the divisor is $N-h,$ 
\item $C^{**},$  the standard estimator, Estimator~\ref{def:empSpatialCorrelation} with the constant normalising factor and the total number of sampled locations $N,$ %the standard estimator with the constant divisor, $N^{-1}$, 
\item $C^{(a)}$, corrected version of $C^{**}$, using the correction method of Estimator~\ref{eq:yaglom_3_2}, 
\item $\widetilde{C}$, Hall's estimator, Estimator~\ref{eq:hall} ,
\item $\widehat{C}_{Q},$ the quantile-based estimator, Estimator~\ref{eq:genton_est},  
\item $\widehat{C}_{N}^{a},$ the tapered estimator, Estimator~\ref{eq:tapered},
\item $\widehat{C}^{B}$, the splines estimator, Estimator~\ref{eq:splines_est}. In this case
 $C^{*}$ was used as the estimated covariance function during the fitting process with  $m=2$ and $p=3.$
\end{itemize}

\begin{figure}[!htb]
  \begin{subfigure}{0.45\textwidth}
    \centering
    \includegraphics[trim= 9mm 0 0 0 ,clip, width=1\linewidth]{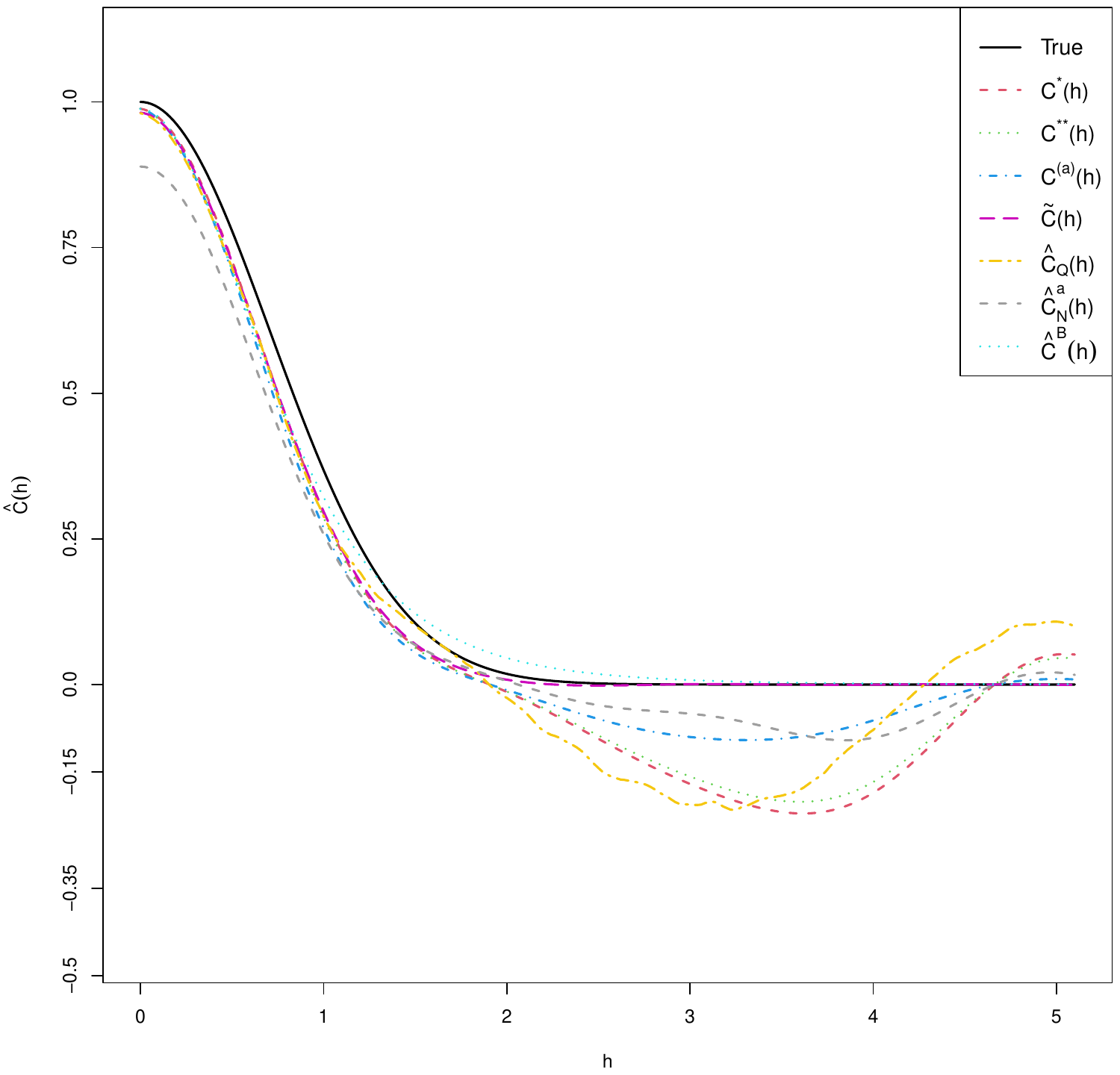}
    \caption{Estimated autocovariance functions over a short range}
    \label{fig:1d_gauss_covariance}
  \end{subfigure}
  \hspace*{0.5cm}
  \begin{subfigure}{0.45\textwidth}
      \centering
      \includegraphics[trim= 9mm 0 0 0 ,clip,width=1\linewidth]{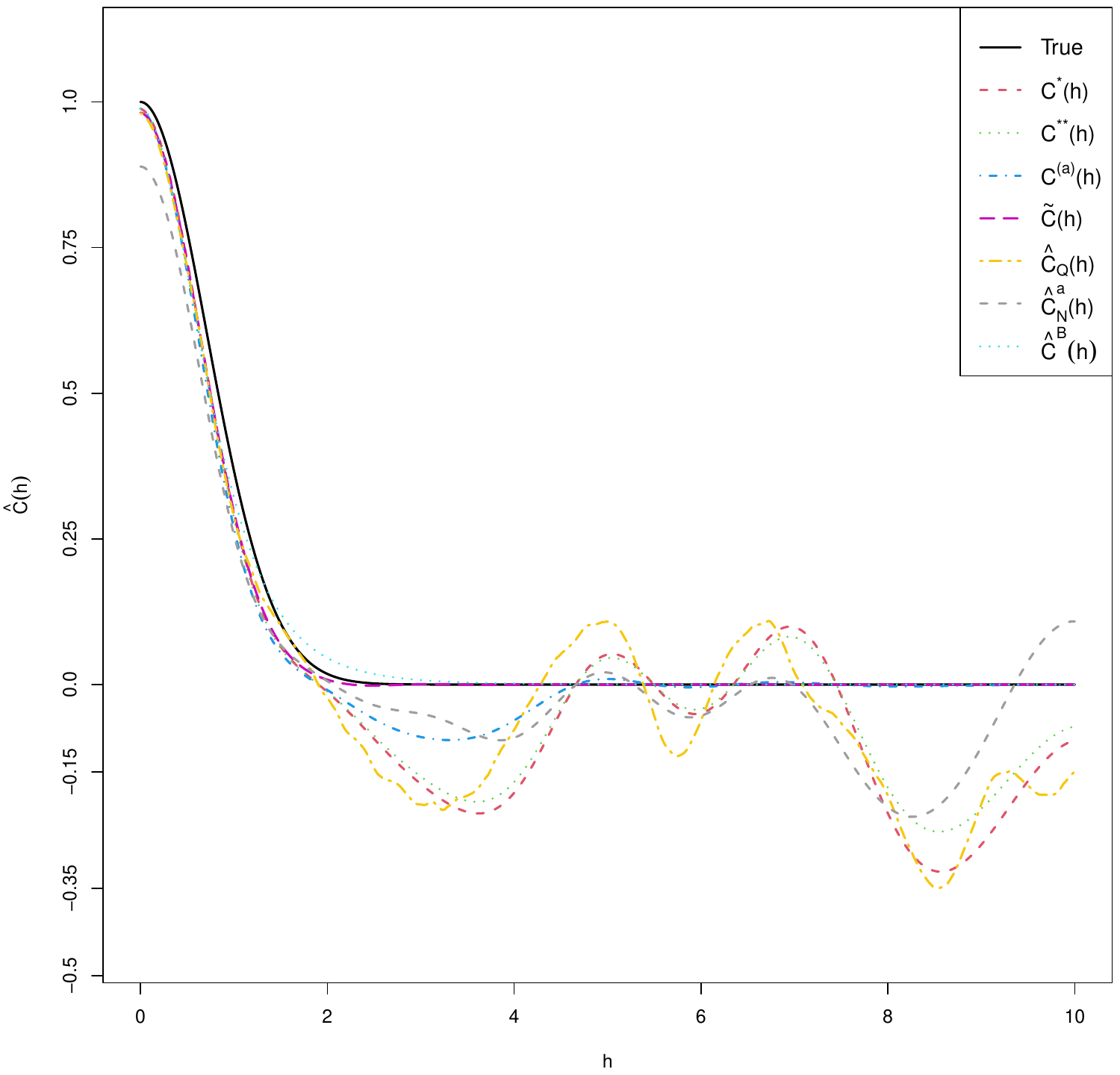}
      \caption{Estimated autocovariance functions over a longer range with prominent waves}
      \label{fig:1d_gauss_waves}
    \end{subfigure}
  \caption{Waves worsening in autocovariance estimators}
\end{figure}

As the spatial case is the focus of this paper, we will briefly demonstrate only one example from the one-dimensional case that highlights the worsening of waves with increasing estimation lag. Let $d=1$ and consider a Gaussian process with the Gaussian covariance and with $\sigma=1.$ Realisations of the process were simulated on the grid $\{0, 0.02, 0.04, \dots, 40\}$ and used to estimate its covariance function.  It is clear from Figure~\ref{fig:1d_gauss_covariance} that all methods estimate the variance of the process well, all around 1, with the lowest being $\widehat{C}_{N}^{a}$ (the grey line) with a value of $0.89$, and the highest being $C^{*}, C^{**},$ and $C^{(a)}$ at $0.99.$ The truncation points for Hall's estimator were chosen as $T_{1}=1.5, T_{2}=2$, and the kernel was the Gaussian kernel, which $C^{(a)}$ also used.
For larger distances, as Hall's estimator brings down the estimate linearly, it removes waves, as one may expect, although other estimators still possess these waves. The waves of estimator~$C^{(a)}$ have small amplitude due to the introduction of a multiplicative kernel transformation.
When the estimation range increases, the amplitudes of the waves become more significant as seen in Figure~\ref{fig:1d_gauss_waves}. From this, one can conclude that most estimators suffer from the issue with waves, except Hall's estimator, $C^{(a)},$ and $\widehat{C}^{B}.$ It is also worth noting the waves occur in similar locations.

In practice, estimating over long distances is not advisable. For example, for the one-dimensional case the range of 10-20\% of the total observation range $N$ is recommended \mbox{\cite[p.~237]{Yaglom1987_1}.} In Figure~\ref{fig:1d_gauss_waves}, 25\% of $N$ was used, whereas Figure~\ref{fig:1d_gauss_covariance} considers only 12.8\%.
Another reason why one should avoid high lag values when estimating the covariance function when using estimator~$C^{*}$, the averaging interval, $N - h$, becomes smaller \cite[p.~237]{Yaglom1987_1}. One may also want to avoid high lag values as the sum of the estimated values of $C^{**}$ is always constant, see the discussion in (iii) from Section~\ref{sec_prop}.

Now, we compare the performance of the considered estimators of the three mentioned isotropic covariance functions using simulated realisations in $\mathbb{R}^2.$ For a fixed  $\tau_{0} \in \R,$ which represents the maximum estimation lag, we use the following five metrics:

\begin{itemize}
    \item 
the {\it area between} the theoretical covariance function and an estimated covariance function,
i.e.
$$
A \lrb{ C(\tau), \widehat{C}(\tau) } = \int_{0}^{\tau_{0}} \abs{ C(\tau) - \widehat{C}(\tau) } \, \text{d}\tau,
$$

\item 
the  {\it maximum vertical distance} between the theoretical covariance function and the estimated covariance function,
$$
D \lrb{ C(\tau), \widehat{C}(\tau) } = \max_{\tau \in [0, \tau_{0}] } \,  \abs{ C(\tau) - \widehat{C}(\tau) } ,
$$

\item the \textit{mean-square prediction error (MSPE)} is considered by taking the differences between the actual values and those computed by kriging. There are two versions, one using the theoretical and estimated covariance function, and one just the estimated covariance function, called \textit{MSPE~gstat} and \textit{MSPE}, respectively.

\item the {\it MSPE~gstat,} requires using both the theoretical and estimated covariance functions, whose steps are outlined below.

%It is a version of the {\it MSPE} metric, which is determined by differences between the actual values and those computed by kriging, as outlined below. The are two versions of this metric. The first, MSPE~gstat, uses the following steps: 

\begin{enumerate}
\item[(1)] Generate a Gaussian random field $X$ on $[-15, 15]^2$ using a grid covering this square with a 0.1 step in both the $x$ and $y$ directions.
\item[(2)] Estimate the covariance functions using the generated values from the subregion $[-10, 10]^2$.
\item[(3)] Estimate the parameter of the theoretical covariance model by using nonlinear least squares. For Gaussian covariance, this is $\sigma$, for Bessel, $\nu$, and for Cauchy, $\gamma$.
\item[(4)] Sample $50$ locations $\bm{t}_{s1}, \dots, \bm{t}_{s50}$ outside the subregion $[-10, 10]^2$.
\item[(5)] Perform kriging on the set of sample locations using the observations from the subregion $[-10, 10]^2$ and obtain $\widehat{X}_{S}$.
\item[(6)] Compute 
$$
\text{MSPE}_g = \frac{1}{50} \sum_{i = 1}^{50} \lrb{ X(\bm{t}_{si}) - \widehat{X}_{S}(\bm{t}_{si}) }^{2}.
$$
\end{enumerate}
To implement the MSPE~gstat, the \textbf{\texttt{R}} package \textbf{\texttt{gstat}} \citep*{gstat2004, gstat2016} was used. As \textbf{\texttt{gstat}} does not support the Bessel and Cauchy covariance functions, the corresponding package functions were modified. 
\item the {\it MSPE,} which is the second version of the MSPE metric, uses only the estimated covariance function  $\widehat{C}.$ For this metric, one uses steps (1), (2) and (4) as above and then proceeds as follows. \begin{enumerate}
\item[(5)] For the $M$ different points $\bm{t}_j,$ construct a $(M + 1) \times (M + 1)$ matrix with the estimated covariances 
$$
\bm{\Gamma} = 
\begin{bmatrix}
\widehat{C}(\norm{\bm{t}_{1} - \bm{t}_{1}}) & \widehat{C}(\norm{\bm{t}_{1} - \bm{t}_{2}}) & \dots  & \widehat{C}(\norm{\bm{t}_{1} - \bm{t}_{M}}) & 1\\
\widehat{C}(\norm{\bm{t}_{2} - \bm{t}_{1}}) & \widehat{C}(\norm{\bm{t}_{2} - \bm{t}_{2}}) & \dots  & \widehat{C}(\norm{\bm{t}_{2} - \bm{t}_{M}}) & 1\\
\vdots                                      & \vdots                                      & \ddots & \vdots                                      & 1 \\
\widehat{C}(\norm{\bm{t}_{M} - \bm{t}_{1}}) & \widehat{C}(\norm{\bm{t}_{M} - \bm{t}_{2}}) & \dots  & \widehat{C}(\norm{\bm{t}_{M} - \bm{t}_{M}}) & 1 \\
1                                           & 1                                           & \dots  & 1                                           & 0 \\
\end{bmatrix} .
$$
To speed computations, we considered the $M=512$ closest locations to each~$\bm{t}_{si}.$ There may not always be 512 different closest locations available, so the largest possible $M$ was used in such cases.

It is worth noting that not all distances between some data locations equate to a sample distance lag, so interpolation can be done to estimate the covariance function at a specific distance. For example, in the simulation studies, the mean covariance value of points within a small neighbourhood can be taken for lags which do not have an empirical covariance value.
\item[(6)] For each randomly sampled location, $\bm{t}_{si}$, compute the distances between it and the $M$ closest sample locations $\bm{t}_j$ from the above step.
\item[(7)] Construct the column vector
$\bm{c} = (\widehat{C}(\norm{\bm{t}_{si} - \bm{t}_{1}}) , \widehat{C}(\norm{\bm{t}_{si} - \bm{t}_{2}})  , \dots , \widehat{C}(\norm{\bm{t}_{si} - \bm{t}_{M}}, 1  )^{\prime}$.
\item[(8)] Compute $\widehat{X}(\bm{t}_{si}) = \lrb{ \lrb{ \bm{\Gamma}^{+} \bm{c} }_{-(M+1)} }^{\prime} (X(\bm{t}_{1}), \dots X(\bm{t}_{M}))^{\prime}$, where $\bm{\Gamma}^{+}$ is the Moore-Penrose inverse, and the notation $\bm{a}_{-j}$ is the vector $\bm{a}$ with the $j^{\text{th}}$ element removed. The Moore-Penrose inverse was chosen as in some cases, due to a computational issue, the matrix~$\bm{\Gamma}$ may be singular.
\item[(9)] Compute the MSPE metric
$$
\text{MSPE} = \frac{1}{50} \sum_{i = 1}^{50} \lrb{ X(\bm{t}_{si}) - \widehat{X}(\bm{t}_{si}) }^{2}.
$$
\end{enumerate}

\item {\it the spectral norm} of a symmetric matrix constructed from the difference between the true covariance function and the empirical covariance function, $D(\tau) = C(\tau) - \widehat{C}(\tau)$ for all lags $\{ 0, \tau_{1}, \dots, \tau_{N} \}$ used in the estimation process.
The matrix $\bm{D}$ is given by
$$
\bm{D} = 
\begin{bmatrix}
D(0)            & D(\tau_{1})     & \cdots & D(\tau_{N - 1}) & D(\tau_{N})     \\
D(\tau_{1})     & D(0)            & \cdots & D(\tau_{N - 2}) & D(\tau_{N - 1}) \\
\vdots          & \vdots          & \ddots & \vdots          & \vdots       \\
D(\tau_{N - 1}) & D(\tau_{N - 2}) & \cdots & D(0)            & D(\tau_{1})     \\
D(\tau_{N})     & D(\tau_{N - 1}) & \cdots & D(\tau_{1})     & D(0)         \\
\end{bmatrix} .
$$
For the two-dimensional case, the spectral norm of the matrix $\bm{D}$  equals its largest singular value.
In the following text, this metric's name will be shortened to \textit{SN}.
\end{itemize}

We computed metrics for 20 realisations of 2D random fields, the average and the average rank of the metrics. The rank was computed for each metric per realisation, resulting in the ranks from 1 to 7 per realisation, where a lower rank indicates better performance.

First, we considered the isotropic Gaussian covariance case, with $\sigma=1.$ The truncation points for Hall's estimator were $T_{1}=1.5,$ $T_{2}=2$, and the kernel was the Gaussian kernel, which was also chosen for $C^{(a)}$.  The estimated Gaussian covariance on the interval $[0, 19.9]$ (i.e. $\tau_0=19.9$) was used for all metrics. The same interval was used for the other covariance model cases too. 

\begin{figure}[!htb]
\centering
  \begin{subfigure}{0.48\textwidth}
	\includegraphics[trim= 9mm 0 0 0, clip, width=\textwidth, height=0.8\textwidth]{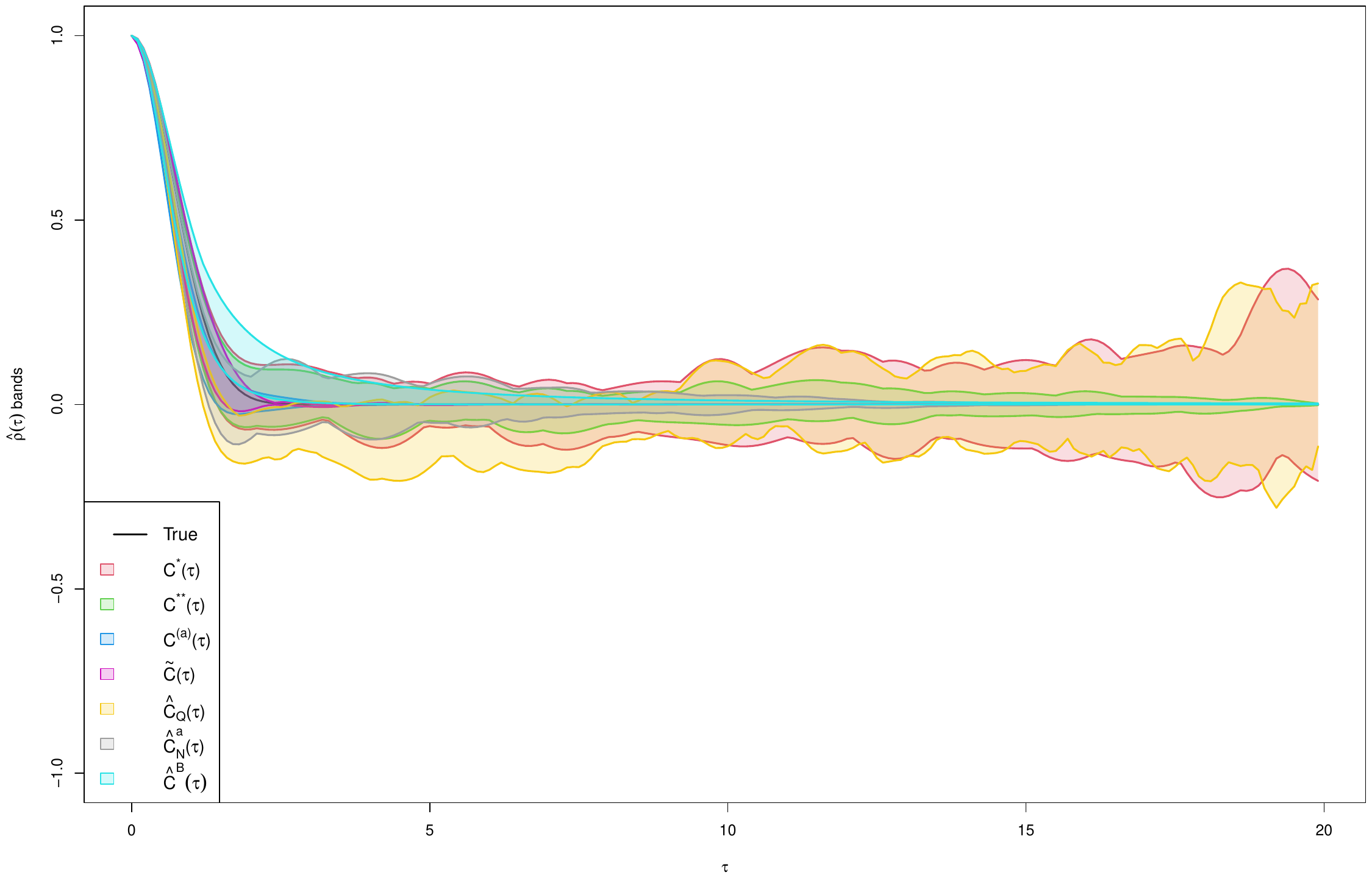}
	\caption{Isotropic Gaussian covariance case}
	\label{fig:2d_iso_gauss}
  \end{subfigure}\hfill
  \begin{subfigure}{0.48\textwidth}
	\centering
	\includegraphics[trim= 9mm 0 0 0, clip, width=\textwidth, height=0.8\textwidth]{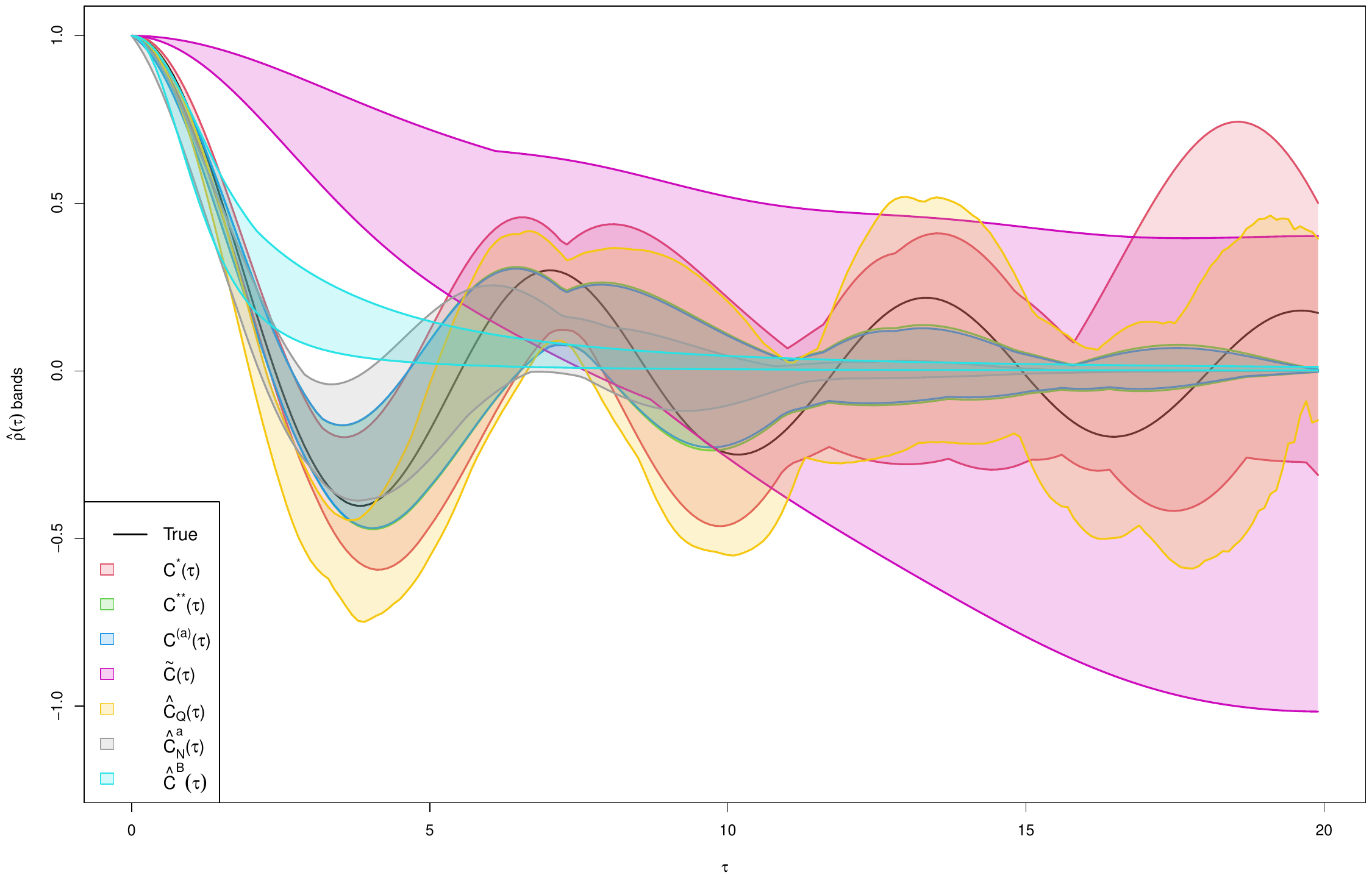}
	\caption{Isotropic Bessel covariance case}
	\label{fig:2d_iso_bessel}
  \end{subfigure}\par
  \vskip\floatsep
  \begin{subfigure}{0.48\textwidth}
  \includegraphics[trim= 9mm 0 0 0, clip, width=\textwidth, height=0.8\textwidth]{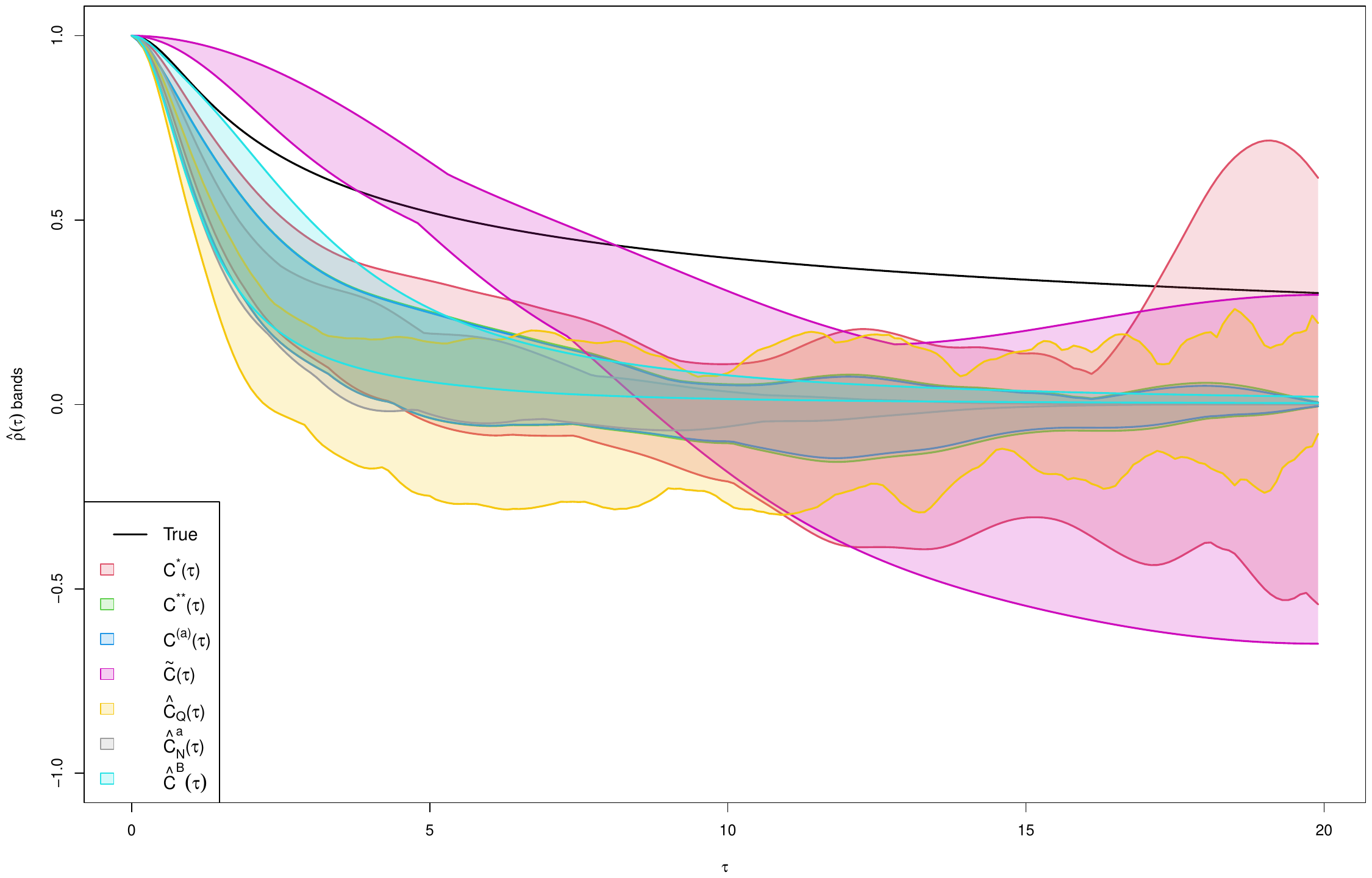}
  \caption{Isotropic Cauchy covariance case}
  \label{fig:2d_iso_cauchy}
  \end{subfigure}
  \caption{Upper and lower bounds for each estimator for 20 realisations}
\end{figure}

Hall's estimator performed the best across area, distance, and SN, whilst the second-best estimator varied across these.
For area and SN, $C^{(a)}$ was the second best estimator, and for distance, it was $C^{**}$, which is reflected partially in the average rank  Table~\ref{tab:different_ranks_gaussian_spatial}.
The large areas of $\widehat{C}_{Q}$ and $C^{*}$ can be seen, especially at the end of the estimation range (see Figure~\ref{fig:2d_iso_gauss}).
These estimators suffered from waves, although Hall's, $C^{(a)}$, and $\widehat{C}^{B}$ reduced the significance of the waves due to their properties.
In this case, $C^{(a)}$ had the lowest MSPE. The MSPE gstat values were quite close, except for $\widehat{C}^{B}$. The ranks in Table~\ref{tab:different_ranks_gaussian_spatial} show that Hall's estimator had the lowest average rank for area, distance, and SN, whilst $\widehat{C}_{Q}$ had the highest average rank for those three. %For MSPE and, except for $C^{(a)}$ and Hall's, all are similar. For MSPE gstat, the ranks are more varied.
For MSPE, $C^{**}$ performs poorly, as do Hall's estimator and $\widehat{C}_{N}^{a}$.
\begin{table}[!htb]
\begin{subtable}{1\textwidth}
	\centering
    \begin{tabular}{|c|c|c|c|c|c|} \hline
		 \diagbox[width=3cm]{Method}{Metric}  & Area & Distance & SN & MSPE & MSPE gstat \\ \hline
        $C^{*}$ & 0.9984164 & 0.1868498 & 5.125285 & 3.810552 & 0.4981920 \\ \hline
        $C^{**}$ & 0.4078536 & 0.0714691 & 3.186527 & 27.400889 & 0.4765476 \\ \hline
        $C^{(a)}$ & 0.1410914 & 0.1183420 & 2.671193 & 2.045802 & 0.4719200 \\ \hline
        $\widetilde{C}$ & 0.0689988 & 0.0481300 & 1.199598 & 10.457498 & 0.4640216 \\ \hline
        $\widehat{C}_{Q}$ & 1.2938920 & 0.2064251 & 11.070654 & 9.062758 & 0.4952457 \\ \hline
        $\widehat{C}_{N}^{a}$ & 0.2839754 & 0.0793276 & 3.172531 & 9.695051 & 0.4607385 \\ \hline
        $\widehat{C}^{B}$ & 0.1748706 & 0.0864897 & 2.928384 & 2.587638 & 0.6607006 \\ \hline
    \end{tabular}
	\caption[Sampled average errors for various estimators of isotropic Gaussian  covariogram]{Sampled average errors for various estimators of isotropic Gaussian  covariogram}
    \label{tab:different_errors_gaussian_spatial}
\end{subtable}

\bigskip
\begin{subtable}{1\textwidth}
	\centering
    \begin{tabular}{|c|c|c|c|c|c|} \hline
		\diagbox[width=4cm]{Method}{Metric's Rank}  & Area & Distance & SN & MSPE & MSPE gstat \\ \hline
        $C^{*}$ & 6.05 & 6.25 & 5.90 & 4.35 & 3.00 \\ \hline
        $C^{**}$ & 4.95 & 2.80 & 3.70 & 4.10 & 3.40 \\ \hline
        $C^{(a)}$ & 2.50 & 4.50 & 3.25 & 2.85 & 4.85 \\ \hline
        $\widetilde{C}$ & 1.15 & 1.35 & 1.30 & 3.30 & 3.35 \\ \hline
        $\widehat{C}_{Q}$ & 6.95 & 6.70 & 6.85 & 4.25 & 6.35 \\ \hline
        $\widehat{C}_{N}^{a}$ & 3.90 & 3.05 & 4.00 & 4.80 & 2.80 \\ \hline
        $\widehat{C}^{B}$ & 2.50 & 3.35 & 3.00 & 4.35 & 4.25 \\ \hline
    \end{tabular}
    \caption[Average performance rank for various estimators of isotropic Gaussian covariogram]{Average performance rank for various estimators of isotropic Gaussian  covariogram}
    \label{tab:different_ranks_gaussian_spatial}
\end{subtable}
\caption[Average performance of the estimators for the isotropic Gaussian case]{Average performance of the estimators for the isotropic Gaussian case}
\end{table}

For simulations from the model with the Bessel covariance function, the parameter $\nu=0$ was selected, which is valid for the considered case $d=2.$ The kernel for Hall's estimator was Gaussian, whilst the wave kernel was chosen for $C^{(a)}$. The truncation points $T_1$ and $T_2$ for Hall's estimator were beyond the used estimation range, making the truncation of the estimator unnecessary. 

Hall's estimator performed the most poorly in this case. It is clear from Figure~\ref{fig:2d_iso_bessel} that Hall's estimator may suffer if there is an early sample spectral frequency with a corresponding negative value. 
%Despite $\widehat{C}^{B}$ having the lowest average error (and thus highest rank), it is clear from Figure~\ref{fig:2d_iso_bessel} that it does not have the periodicity of other estimators, such as $C^{*}$ or $\widehat{C}_{Q}$. This can be problematic when performing estimation as the estimated function does not act like the true function, rather a short-memory function. The same issue occurs with $C^{(a)}$, although one may wish to adjust the point where the kernel function reaches zero if the estimation is to be extended. ? Where did I find this information? I think I read the wrong table.
$C^{**}$ had the lowest error for area and SN, $C^{(a)}$ for distance, $\widehat{C}^{B}$ for MSPE, and for MSPE gstat, all were similar with the exception of $\widehat{C}^{B}$ (see Table~\ref{tab:different_errors_bessel_spatial}). $\widehat{C}^{B}$ having the lowest MSPE was quite surprising, given it did not follow the shape of the true covariance function, and had the second largest area. Despite $C^{(a)}$ and $C^{**}$ having the lowest errors, it is visually clear that neither followed the true function as nice as $C^{*}$ or $\widehat{C}_{Q}$ did. $C^{**}$ and $\widehat{C}^{B}$ were going to zero, whilst $C^{(a)}$ decayed to zero due to the use of a kernel function, and $C^{*}$ and $\widehat{C}_{Q}$ roughly followed the true function, despite having a wide prediction range.
If ranks are considered (see Table~\ref{tab:different_ranks_bessel_spatial}), $C^{**}$ was still the best for area and SN, whilst $C^{(a)}$ for distance, $\widehat{C}_{N}^{a}$ for MSPE, as discussed above. This is surprising given they did not follow the shape of the function nicely.
    
\begin{table}[!htb]
\begin{subtable}{1\textwidth}
	\centering
    \begin{tabular}{|c|c|c|c|c|c|} \hline
		\diagbox[width=3cm]{Method}{Metric}  & Area & Distance & SN & MSPE & MSPE gstat \\ \hline
        $C^{*}$ & 2.489791 & 0.3507251 & 15.61262 & 125.475762 & 0.0056134 \\ \hline
        $C^{**}$ & 2.098724 & 0.2503224 & 12.70419 & 6.760378 & 0.0084915 \\ \hline
        $C^{(a)}$ & 2.109226 & 0.2496318 & 12.78919 & 106.180658 & 0.0086290 \\ \hline
        $\widetilde{C}$ & 7.513908 & 1.0682663 & 82.24297 & 75.854967 & 0.0764184 \\ \hline
        $\widehat{C}_{Q}$ & 2.748823 & 0.3275784 & 18.99515 & 443.374390 & 0.0049442 \\ \hline
        $\widehat{C}_{N}^{a}$ & 2.492002 & 0.2714209 & 16.79564 & 2.626784 & 0.0140113 \\ \hline
        $\widehat{C}^{B}$ & 3.486087 & 0.5350350 & 29.29765 & 1.950848 & 0.1618823 \\ \hline
    \end{tabular}
	\caption[Sampled average errors for various estimators of isotropic Bessel  covariogram]{Sampled average errors for various estimators of isotropic Bessel  covariogram}
    \label{tab:different_errors_bessel_spatial}
\end{subtable}

\bigskip
\begin{subtable}{1\textwidth}
	\centering
    \begin{tabular}{|c|c|c|c|c|c|} \hline
		\diagbox[width=4cm]{Method}{Metric's Rank}  & Area & Distance & SN & MSPE & MSPE gstat \\ \hline
        $C^{*}$ & 3.20 & 4.00 & 2.80 & 5.20 & 1.875 \\ \hline
        $C^{**}$ & 1.90 & 2.20 & 1.75 & 3.95 & 3.200 \\ \hline
        $C^{(a)}$ & 2.40 & 2.10 & 2.25 & 4.00 & 3.800 \\ \hline
        $\widetilde{C}$ & 7.00 & 7.00 & 7.00 & 5.35 & 6.100 \\ \hline
        $\widehat{C}_{Q}$ & 4.15 & 3.85 & 4.35 & 5.30 & 1.225 \\ \hline
        $\widehat{C}_{N}^{a}$ & 3.75 & 3.00 & 4.00 & 1.95 & 4.900 \\ \hline
        $\widehat{C}^{B}$ & 5.60 & 5.85 & 5.85 & 2.25 & 6.900 \\ \hline
    \end{tabular}
    \caption[Average performance rank for various estimators of isotropic Bessel covariogram]{Average performance rank for various estimators of isotropic Bessel covariogram}
    \label{tab:different_ranks_bessel_spatial}
\end{subtable}
\caption[Tables of averages for the case of isotropic Bessel covariance]{Average performance of the estimators for the isotropic Bessel case}
\end{table}

Finally, for the Cauchy covariance function case, $\gamma=0.2$ was selected.  For Hall's estimator and $C^{(a)}$, the rational quadratic kernel was chosen, and the truncation points were once again beyond the estimation range, making truncation unnecessary.

In this case, for lower values of $\tau$, the estimators performed well, but as the estimation range increased, the estimators performed worse, which can be seen in Figure~\ref{fig:2d_iso_cauchy}. In terms of error, Hall's estimator performed the best for area and SN, whilst $\widehat{C}^{B}$ for distance, $C^{*}$ for MSPE, and for MSPE gstat, $\widehat{C}_{Q}$ was the best, and all others were similar.
If we considered the average rank instead (see Table~\ref{tab:different_ranks_cauchy_spatial}), it is clear that $\widehat{C}^{B}$ had the lowest average rank for area and distance, and Hall's for SN, but it had the worst MSPE rank, and $\widehat{C}_{Q}$ had the best MSPE gstat rank.
Unlike Hall's estimator, $\widehat{C}^{B}$ did not flare out after $\tau=10$, and instead, the minimum and maximum values of the band approached one another. As a result, $\widehat{C}^{B}$ can be seen as a more stable estimator when dealing with noncyclic nonnegative covariance functions.

\begin{table}[!htb]
\begin{subtable}{1\textwidth}
	\centering
    \begin{tabular}{|c|c|c|c|c|c|} \hline
		\diagbox[width=3cm]{Method}{Metric}  & Area & Distance & SN & MSPE & MSPE gstat \\ \hline
        $C^{*}$ & 7.486252 & 0.6165021 & 70.18825 & 2.582530 & 7.051711 \\ \hline
        $C^{**}$ & 7.191006 & 0.4875699 & 72.18749 & 3.678911 & 6.974136 \\ \hline
        $C^{(a)}$ & 7.187045 & 0.4864138 & 72.23088 & 5.487214 & 6.970320 \\ \hline
        $\widetilde{C}$ & 5.671946 & 0.5083464 & 36.96571 & 786.580629 & 7.069838 \\ \hline
        $\widehat{C}_{Q}$ & 8.393729 & 0.6678209 & 91.26490 & 10.102567 & 5.850791 \\ \hline
        $\widehat{C}_{N}^{a}$ & 7.212647 & 0.4913294 & 73.58152 & 17.153026 & 7.003175 \\ \hline
        $\widehat{C}^{B}$ & 6.280143 & 0.3943084 & 61.03689 & 5.872001 & 6.990817 \\ \hline
    \end{tabular}
	\caption[Sampled average errors for various estimators of isotropic Cauchy  covariogram]{Sampled average errors for various estimators of isotropic Cauchy covariogram}
    \label{tab:different_errors_cauchy_spatial}
\end{subtable}

\bigskip
\begin{subtable}{1\textwidth}
	\centering
    \begin{tabular}{|c|c|c|c|c|c|} \hline
		\diagbox[width=4cm]{Method}{Metric's Rank}  & Area & Distance & SN & MSPE & MSPE gstat \\ \hline
        $C^{*}$ & 4.80 & 5.65 & 3.70 & 3.50 & 4.3 \\ \hline
        $C^{**}$ & 4.35 & 3.60 & 4.40 & 3.20 & 4.2 \\ \hline
        $C^{(a)}$ & 3.95 & 3.65 & 4.75 & 3.25 & 4.0 \\ \hline
        $\widetilde{C}$ & 2.25 & 3.55 & 1.10 & 6.95 & 4.2 \\ \hline
        $\widehat{C}_{Q}$ & 6.40 & 6.30 & 6.80 & 3.70 & 2.6 \\ \hline
        $\widehat{C}_{N}^{a}$ & 4.35 & 3.80 & 5.10 & 3.25 & 4.4 \\ \hline
        $\widehat{C}^{B}$ & 1.90 & 1.45 & 2.15 & 4.15 & 4.3 \\ \hline
    \end{tabular}
    \caption[Sampled average errors for various estimators of isotropic Cauchy  covariogram]{Sampled average errors for various estimators of isotropic Cauchy  covariogram}
    \label{tab:different_ranks_cauchy_spatial}
\end{subtable}
\caption[Tables of averages for the case of isotropic Cauchy covariance]{Average performance of the estimators for the isotropic Cauchy case}
\end{table}

\section{Discussion and conclusion}\label{sec12}
Several nonparametric methods for estimating spatial covariance functions were reviewed, studied, and compared based on unified principles. The main focus was on the case of a single realisation and the methods utilising the actual values of observations. Some surprising drawbacks of several well-known estimators were identified. Numerical studies compared the accuracy of the considered estimators using various metrics and three theoretical covariogram models. When dealing with random fields whose covariance functions are not oscillating and decay sufficiently fast, such as the Gaussian covariance, estimators like Hall's estimator and $C^{(a)}$ outperform other estimators. However, in the case of cyclic or long-range dependence, the vanishing estimators often fail to capture the behaviour of the theoretical covariogram model. In such instances, other estimators, such as $C^{*}$ in the Bessel case, may be better suited, even if they have some constant summability properties. The results are useful for understanding the limitations and proper usage of covariograms in various applications, including kriging, monitoring network optimisation, and cross-validation.

Future work includes an overview and comparative studies of other estimation methods for covariograms, as well as similar studies for estimation methods of spectral densities. Other potential extensions could involve analogous investigations of estimates of dependency and the spectrum of multivariate and spherical spatial data, as well as spatio-temporal data.

\section*{Acknowledgments}
This research was partially supported under the Australian Research Council's Discovery Projects funding scheme (project number  DP220101680). Andriy Olenko was also partially supported by La Trobe University's SCEMS CaRE and Beyond grant. We would like to thank Professors N.Cressie, N.Leonenko, and E.Porcu for their constructive feedback on the early draft of this paper.

\bibliography{bib}

\appendix

\section{R and Python packages with estimators of covariance functions}\label{appRpackages}
\proglang{R} packages: \texttt{stats, RandomFields, spatcov, fields, TSA, astsa, forecast, Stat2Data}

\proglang{Python} packages: \texttt{NumpPy, statsmodels, pandas}

Many of these packages simply use wrapper functions for the \texttt{acf} function.
It is important to note that this list is not exhaustive.

%The format of the list will be \texttt{package: function}. Many of these functions are simply wrapper functions for the stats' \texttt{acf} function.
%\begin{itemize}
%    \item \texttt{stats: acf}
%    \item \texttt{RandomFields: RFcov}
%    \item \texttt{spatcov: spatcov}
%    \item \texttt{fields: vgram(..., type='covariogram')}
%    \item \texttt{TSA: acf}
%    \item \texttt{astsa: acf1}
%    \item \texttt{forecast: Acf}
%    \item \texttt{Stat2Data: sluacf}
%\end{itemize}
%
%In Python
%\begin{itemize}
%    \item \texttt{NumpPy: correlate}
%    \item \texttt{statsmodels: acf}
%    \item \texttt{pandas: autocorr}
%\end{itemize}

\section{Non-positive-definiteness of classical estimator}\label{appNonpositive}

Consider the points $\bm{s}_{1} = (1, 0)$, $\bm{s}_{2}=(2, 0)$ and $\bm{s}_{3}=(3, 0)$, with values $X(\bm{s}_{1}) = X(\bm{s}_{3}) = 1$, $X(\bm{s}_{2}) = 0$, and coefficients $a_{1}=\iu$, $a_{2} = 1$, and $a_{3}=-\iu$.
	We will have three pairs of points for the distance $\tau=0$, $\{ (\bm{s}_{1}, \bm{s}_{1}) , (\bm{s}_{2}, \bm{s}_{2}) , (\bm{s}_{3}, \bm{s}_{3}) \}$,
	at $\tau=1$ there are four pairs, $\{ (\bm{s}_{1}, \bm{s}_{2}) , (\bm{s}_{2}, \bm{s}_{1}) , (\bm{s}_{2}, \bm{s}_{3}), (\bm{s}_{3}, \bm{s}_{2}) \}$, and for $\tau=2$ there are two pairs of points,
	$\{ (\bm{s}_{1}, \bm{s}_{3}) , (\bm{s}_{3}, \bm{s}_{1}) \}$. %$\abs{a_{1}}^{2} = \abs{a_{2}}^{2} = \abs{a_{3}}^{2} = 1$,
	%$a_{1}\mean{a}_{2} = a_{2}\mean{a}_{3} = i$, $a_{1}\mean{a}_{3} = a_{3}\mean{a}_{1} = -1$, $a_{2}\mean{a}_{1} = a_{3}\mean{a}_{2} = -i$.
	Expanding the sum, we obtain a contradiction with the positive-definite condition:
	\[\sum_{i=1}^3\sum_{j=1}^3a_jX(s_i)\bar{X}(s_j)\bar{a}_j= \frac{1}{3} \cdot  2 + \frac{1}{4} \cdot 0  + \frac{1}{2} \cdot -2 = -\frac{1}{3}
		< 0.
	\]

\section{Pseudocode for computing isotropic covariance function}\label{appPseudocode}
\begin{algorithm}
	\KwData{$N$ sample values of a random field $X$, desired distance $\tau$}
	sum $\gets 0$\;
	$N(\tau) \gets \emptyset$\;
	\For{Each coordinate $\bm{t}_{i}$ in $X$}{
		\For{Each coordinate $\bm{t}_{j}$ in $X$}{
			\If{$\norm{\bm{t}_{i} - \bm{t}_{j}} = \tau$}{
				Append $(\bm{t}_{i}, \bm{t}_{j})$ to $N(\tau)$\;
			}
		}
	}
	\For{Each pair  $(\bm{t}_{i}, \bm{t}_{j})$  in $N(\tau)$}{
		sum $\gets$ sum $ + X(\bm{t}_{i}) X(\bm{t}_{j})$\;
	}
	\Return $\lrb{ \text{sum} / \abs{N(\tau)} }$
	
	\KwResult{An estimated isotropic covariance function for $X$ at a desired radius~$\tau$}
 \end{algorithm}

\section[Positive-definiteness of classical estimator with constant N(h)]{Positive-definiteness of classical estimator with constant $N(h)$}\label{appPositive}
Without loss of generality, assume that $X$ is a zero-mean random field.
Let $K = \{ \widetilde{s}_{k} \}$ denote the set of observations, and let us set $X(s) = 0$ if $s \notin K$. As in the covariance function estimator, we use only the values of $X$ over the set $K$, this assumption does not change the estimate but proves to be useful to avoid considering different subcases.
	For $I = \{ s_{i} , i=1, \dots, n \}$, consider a set $S_{K, I} = \{ \widetilde{s}_{k} - s_{i}, \widetilde{s}_{k} \in K , s_{i} \in I \}$, and let 
	$a_{1}, \dots, a_{n} \in \CC$ be arbitrary constants. Then, for  $N(s_{i} - s_{j}) = N,$ where $N$ is constant, we obtain
		\begin{align}
			& \sum_{i,j=1}^{n} a_{i} \widehat{C}(s_{i} - s_{j}) \mean{a}_{j} \nonumber 
			= \sum_{i,j=1}^{n} a_{i} \lrb{ \frac{1}{\abs{N(s_{i} - s_{j})}} \sum_{k \in N(s_{i} - s_{j})} X(\widetilde{s}_{k}) \mean{X(\widetilde{s}_{k} + (s_{i} - s_{j}))} } \mean{a}_{j}  \nonumber \\
			&= \frac{1}{N} \sum_{i,j=1}^{n} a_{i} \lrb{ \sum_{s \in S_{K, I}} X(s + s_{i}) \mean{X(s + s_{j})} } \mean{a}_{j} = \frac{1}{N} \sum_{s \in S_{K, I}} \sum_{i,j=1}^{n} a_{i} X(s + s_{i}) \mean{X(s + s_{j})}  \mean{a}_{j} \nonumber \\
			&= \frac{1}{N} \sum_{s \in S_{K, I}} \left| \sum_{i=1}^{n} a_{i} X(s + s_{i}) \right|^{2} \geq 0 . \nonumber
		\end{align}

\section{Summability identity for classical spatial covariogram estimator}\label{appSummability}

By Definition~(\ref{def:empSpatialCorrelation}), the standard estimate of spatial autocorrelation function is
    \begin{align*}
        \widehat{\rho}(\bm{h}) &= \frac{\sum_{\bm{t}_{i}, \bm{t}_{j} \in N(\bm{h})}(X(\bm{t}_{i}) - \mean{X}) (X(\bm{t}_{j}) - \mean{X})/ N(\bm{h}) }{ \sum_{i=1}^{N}  (X(\bm{t}_{i}) - \mean{X})^{2}/N  } ,
    \end{align*}
   
    Assume that $N(\bm{h}) \equiv N.$ Then,
    \begin{align*}
        \sum_{\bm{h} \in H} \widehat{\rho}(\bm{h}) &= \sum_{\bm{h} \in H} \frac{\sum_{\bm{t}_{i}, \bm{t}_{j} \in N(\bm{h})}(X(\bm{t}_{i}) - \mean{X}) (X(\bm{t}_{j}) - \mean{X}) }{ \sum_{i=1}^{N}  (X(\bm{t}_{i}) - \mean{X})^{2} } \\
        &= \frac{ \sum_{\bm{h} \in H }\sum_{\bm{t}_{i}, \bm{t}_{j} \in N(\bm{h})}(X(\bm{t}_{i}) - \mean{X}) (X(\bm{t}_{j}) - \mean{X}) }{ \left( \sum_{i=1}^{N}  (X(\bm{t}_{i}) - \mean{X}) \right)^{2} - \sum_{i=1}^{N} \sum_{\substack{j = 1 \\ j \neq i}}^{N} (X(\bm{t}_{i}) - \mean{X}) (X(\bm{t}_{j}) - \mean{X}) } \\
        &= \frac{ \sum_{\bm{h} \in H } \sum_{\bm{t}_{i}, \bm{t}_{j} \in N(\bm{h})}(X(\bm{t}_{i}) - \mean{X}) (X(\bm{t}_{j}) - \mean{X}) }{ -\sum_{\bm{h} \in H} \sum_{\bm{t}_{i}, \bm{t}_{j} \in N(\bm{h})} (X(\bm{t}_{i}) - \mean{X}) (X(\bm{t}_{j}) - \mean{X}) } = -1 .
    \end{align*}
    Note, that a similar identity also holds for the case of nonconstant $N(\bm{h}).$ Namely, from the above identity, it follows that
    \begin{align*}
        -1 &= \sum_{\bm{h} \in H} \widehat{\rho}(\bm{h}) = \frac{ \sum_{\bm{h} \in H} \widehat{C}^{**}(\bm{h}) }{\widehat{\text{Var}}(X)}
         = \frac{ \sum_{\bm{h} \in H} \frac{1}{N} \sum_{\bm{t}_{i}, \bm{t}_{j} \in N(\bm{h})} (X(\bm{t}_{i}) - \mean{X}) (X(\bm{t}_{j}) - \mean{X}) } {\frac{1}{N} \sum_{i=1}^{N} (X(\bm{t}_{i}) - \mean{X})^{2} } \\
        &= \sum_{\bm{h} \in H} \frac{\abs{N(\bm{h})}}{N} \frac{1}{\abs{N(\bm{h})}}\frac{ \sum_{\bm{t}_{i}, \bm{t}_{j} \in N(\bm{h})} (X(\bm{t}_{i}) - \mean{X}) (X(\bm{t}_{j}) - \mean{X}) } {\frac{1}{N} \sum_{i=1}^{N} (X(\bm{t}_{i}) - \mean{X})^{2} } 
        = \sum_{\bm{h} \in H} \frac{\abs{N(\bm{h})}}{N} \widetilde{\rho}(\bm{h}),
    \end{align*}    
    where $\widetilde{\rho}(\bm{h}) = \widehat{C}(\bm{h}) / \widehat{C}(\bm{0})$.
    Thus, the weighted sum of $\widetilde{\rho}(\bm{h})$ maintains a constant value regardless of the observations $X(\bm{t}_{i})$, and that the weights $\abs{N(\bm{h})} / N$ depend only the locations of the observations, not the values at the observations.
    
\section{Relation between classical covariogram and semivariogam estimators}\label{appRelation}
 The relation $\widehat\gamma(\bm{h}) = \widehat C(\bm{0}) - \widehat C(\bm{h})$ does not hold for the classical estimators as follows from
	$$
	\begin{aligned}
		\widehat{\gamma}(\bm{h}) &= \frac{ 1 }{ 2\abs{N(\bm{h})} } \sum_{N(\bm{h})} \lrb{  X(\bm{t}_{i}) - X(\bm{t}_{j}) }^{2}
		= \frac{ 1 }{ 2\abs{N(\bm{h})} } \sum_{N(\bm{h})} \lrb{  (X(\bm{t}_{i}) - \mean{X}) - (X(\bm{t}_{j}) - \mean{X}) }^{2} \\
		&= \frac{ 1 }{ 2\abs{N(\bm{h})} } \sum_{i \in N(\bm{h})} 2(X(\bm{t}_{i}) - \mean{X})^{2} - \frac{ 1 }{ 2\abs{N(\bm{h})} } \sum_{N(\bm{h})} 2(X(\bm{t}_{i}) - \mean{X})(X(\bm{t}_{j}) - \mean{X}) \\
		&= \frac{ 1 }{ \abs{N(\bm{h})} } \sum_{i \in N(\bm{h})} (X(\bm{t}_{i}) - \mean{X})^{2} - \widehat{C}(\bm{h}) .
	\end{aligned}
	$$
	 As the estimate $\widehat{C}(\bm{0}) = N^{-1} \sum_{i = 1}^{N} ( X(\bm{t}_{i}) - \mean{X} )^{2}$ incorporates all spatial locations,
	not only those separated by a vector $\bm{h}$, 
the relation $\widehat\gamma(\bm{h}) = \widehat C(\bm{0}) - \widehat C(\bm{h})$ is not valid in general, except the case when $\bm{h} = \bm{0}$.
\end{document}